\documentstyle[preprint,floats,epsf,eqsecnum,aps,tighten]{revtex}

\newcommand{\be}{\begin{equation}}
\newcommand{\ee}{\end{equation}}
\newcommand{\bea}{\begin{eqnarray}}
\newcommand{\eea}{\end{eqnarray}}
\def\s{\ \! / \! \! \! \!}

\def\stack{\stackrel}

\begin{document}
\draft
\preprint{
\parbox{1.5in}{\leftline{JLAB-THY-99-xx}
                 \leftline{WM-99-xx}
                 \leftline{}\leftline{}\leftline{}\leftline{}}}
\title{Quark-Antiquark Bound States in the\\ Relativistic Spectator Formalism}
\author{\c{C}etin \c{S}avkl{\i}$^{1}$ \& Franz Gross$^{1,2}$}
\address{ $^1$Department of Physics, College of William and Mary, Williamsburg,
Virginia 23187\\
$^2$Jefferson Lab, 
12000 Jefferson Avenue, Newport News, VA 23606
}
\date{\today}
\maketitle
\begin{abstract}
The quark-antiquark bound states are discussed using the relativistic spectator
(Gross) equations. A relativistic covariant framework for analyzing
confined bound  states is developed. The relativistic linear 
potential developed in
an  earlier work is proven to give vanishing meson$\rightarrow$
$q+\bar{q}$ decay amplitudes, as required by confinement. The 
regularization of the
singularities in the linear potential that are associated with 
nonzero energy transfers
(i.e. $q^2=0,q^{\mu}\neq0$) is improved.  Quark mass functions that 
build chiral symmetry
into the theory and explain the connection between the current quark 
and constituent quark
masses are introduced.  The formalism is applied to the description 
of pions and kaons
with reasonable results.
\end{abstract}

\section{Introduction}
\label{introduction}

Description of simple hadrons in terms of quark-gluon degrees of freedom
has long been an active area in physics. With the advent of Jefferson 
Laboratory,
which  operates at intermediate energies and therefore probes the structure
of hadrons, there are new opportunities to test simple theoretical
descriptions of quark interactions. The first natural step in this direction
is a thorough understanding of how to treat the relativistic quark-anti quark
bound state problem. In this context, NJL inspired models have gained
popularity in recent years\cite{GROSS,ROBERTS1}. The common goal of these
works is to bridge the gap between nonrelativistic quark models and more
rigorous approaches, such as lattice gauge theory or Feynman-Schwinger
calculations. While the Euclidean metric  based calculations are increasingly
popular, their applicability, because  of the extrapolations involved, is only
limited to light bound states such as the pion and kaon. Therefore, it is
important to develop Minkowski metric based  models which can be used 
over a wider
scale of energies. One such work using the spectator formalism was developed in
Ref.~\cite{GROSS}. In  those works a relativistic generalization of the linear
potential was  developed and the pion was shown to be massless in the 
chiral limit.
However,  the calculations involved some approximations and related conceptual
problems.  In this work we improve and simplify the model presented in those
works and address in detail some of the conceptual issues related to
confinement.

If a quark-antiquark pair (referred to collectively as ``quarks'') is confined
to a meson bound  state with mass $\mu$, then the bound state can not decay into
two free quarks, even if the sum of the quark masses is less than the 
bound state
mass. This trivial statement can be realized by two possible 
mechanisms: either
(a) the  quark propagators are free of timelike mass 
poles,\cite{ROBERTS1} or (b) the
vertex function of the bound state {\it vanishes\/} when both quarks 
are  on-shell.  In
this work we {\it prove\/} that the Gross equation supports the second
mechanism of confinement. The  first mechanism, which is commonly used in
Euclidean metric based  calculations, is a stronger constraint since it forbids
any free quark states. On the other hand, the Gross equation allows one of the
two quarks in a meson to be on-shell, but insures that the matrix element which
couples the bound state to {\it two\/} free quarks vanishes. The spectator
formalism facilitates the  use of the Minkowski metric, and the confinement
mechanism of this approach has a closer resemblance to nonrelativistic
models.

The organization of the paper is as follows: In 
Sec.~\ref{confinement} we review
the formalism for nonrelativistic confinement in momentum space.  In 
Sec.~III we
outline the general philosophy of the spectator approach to the treatment of
confined systems, examine the implications of confinement for the scattering
amplitude, and prove that the relativistic linear potential used in 
earlier works
automatically insures that $\mu\rightarrow q+\bar{q}$ vanishes at the momentum
where decay of the state into two physical quarks would otherwise be 
kinematically
possible.  The treatment is first presented for scalar particles, and then
generalized to fermions.  In Sec.~\ref{chiral_limit} we construct quark mass
functions that have the correct chiral limit and preserve asymptotic 
freedom.  Our
numerical results for pseudoscalar bound states are presented in
Sec.~\ref{results}, and some conclusions are given in Sec.~\ref{conclusion}.

\section{Nonrelativistic confinement in momentum space}
\label{confinement}


We start by reviewing the discussion of confinement
within the context of the nonrelativistic Schr\"odinger equation given
in Ref.~\cite{GROSS}. We will denote potentials in coordinate space by
$\tilde{V}$ and in momentum space by $V$. The nonrelativistic linear 
potential is
\begin{equation}
\tilde{V}(r)=\sigma r\, .
\end{equation}
This potential can be constructed from familiar Yukawa-like potentials in two
different ways:
\bea
\tilde{V}(r)=\lim_{\epsilon\to0}\cases{ \tilde{V}_S(r) \equiv \sigma r
e^{-\epsilon r} & \rm{(a)} \cr
\tilde{V}_L(r) \equiv -{\displaystyle{\frac{\sigma}{\epsilon}}}\left( 
e^{-\epsilon
r} - 1
\right) =\tilde{V}_A(r) + {\displaystyle\frac{\sigma}{\epsilon}}\, . 
\qquad \qquad
& \rm{(b)}} \label{2spots}
\eea
These various potentials are shown in Fig.~1 for the illustrative case of
$\epsilon=0.1$ and $\sigma=0.2$.

\begin{figure} 
\centerline{
    \epsfysize=3.0in
\epsffile{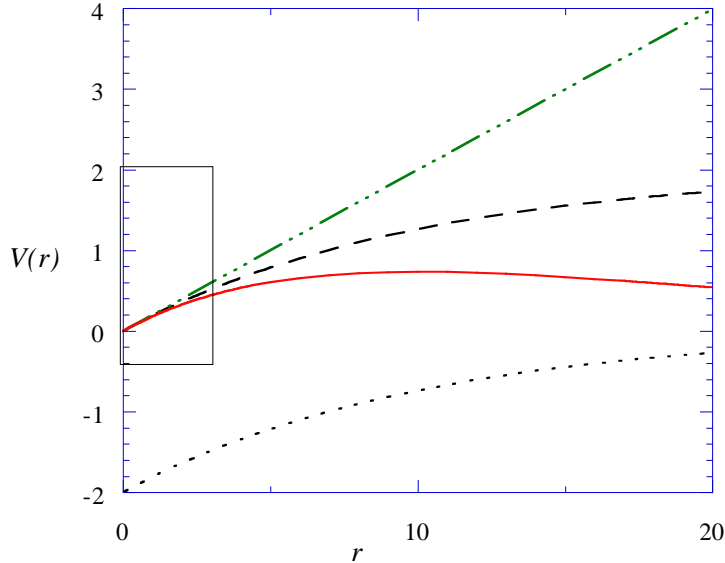}}
\caption{The linear potential in coordinate space for $\epsilon=0.1$ and
$\sigma=0.2$.  The solid line is $\tilde V_S(r)$, the dashed line is $\tilde
V_L(r)$, the dotted line is $\tilde V_A(r)$, and the dot-dashed line is 
$\tilde V(r)$.  For ``small'' $r< 1/\epsilon$ (the region inside the small 
box) $\tilde V_L(r)$ and $\tilde V_S(r)$ are both approximately equal to 
$\sigma r$.}
\label{linear.fig}
\end{figure}

Note that the two potentials $\tilde V_S(r)$ and $\tilde V_L(r)$ both 
approximate the
linear potential $\tilde{V}(r)$ when $r<<1/\epsilon$, but that these 
two approximate
potentials behave very differently at large $r$.  The potential 
$\tilde V_S(r)\to0$ at
large $r$, so that, strictly speaking, it does not confine particles at all.
This potential always permits scattering, although when $\epsilon$ is small the
scattering is strongly resonant, and the wave function is significant 
at small $r$
only for energies near one of the allowed resonances.  The width of 
these resonance
states becomes narrower, and their wave function approaches that of a bound
state, as $\epsilon\to0$.  In contrast, the potential $\tilde 
V_L(r)\to 1/\epsilon$ as
$r\to\infty$ and therefore binds particles with energies $E<1/\epsilon$.  As
$\epsilon\to0$ this potential does not permit scattering; it has a 
spectrum of bound
states only.

Yet for sufficiently small $\epsilon$, it should be possible to move 
freely from one
of these potentials to the other, and the results obtained with 
either form should be
equivalent.  We will return to this later in this section.  Now we follow
Ref.~\cite{GROSS} and work with $\tilde{V_L}$ given in Eq.~(\ref{2spots}b).

The momentum space form of this potential can be written
\begin{eqnarray}
V_L({\bf q})=&&\lim_{\epsilon\rightarrow 0}\;
\left[V_A({\bf q})-\delta^3({q})\int
d^3q'\,V_A({\bf q}')\right] \, ,
\label{nrpotential}
\end{eqnarray}
where
\begin{eqnarray}
V_A({\bf q})= -\frac{8\pi\sigma}{({\bf q}^2+\epsilon^2)^2}\, ,
\label{nrpotential2}
\end{eqnarray}
Note that the second term (the ``subtraction term'') insures that
\be
\int d^3q\; V_L({\bf q})=0 \, , \label{identity}
\ee
which is the momentum space form of the statement that $\tilde V(r=0)=0$.
The Fourier transform of $V_A$ is, for finite $\epsilon$,
\begin{eqnarray}
\tilde V_A({\bf r})&=&\int\frac{d^3q}{(2\pi)^3}\,e^{-iq\cdot r}\,V_A({\bf q})\\
             &=&-\sigma\frac{e^{-\epsilon r}}{\epsilon}\ \ \simeq\ \
\lim_{\epsilon\rightarrow 0} \sigma \left(r-\frac{1}{\epsilon}\right) \, ,
\end{eqnarray}
and the subtraction term cancels the singular $1/\epsilon$ term insuring
that the linear part of the potential has the correct behavior in the limit as
$\epsilon \to0$ and that it vanishes at the origin ($r=0$).  Now, 
adding a constant
potential $V_C$
\bea
V_C(r)=&&-C \nonumber\\
V_C({\bf q})=&&-(2\pi)^3\delta^3({q}) C \, ,
\eea
to the linear potential (\ref{nrpotential}), and inserting the total 
potential into the
momentum space Schr\"odinger equation gives
\begin{equation}
\left[ \frac{{\bf p}^2}{2m_R}-E\right]\Psi({\bf p},p_0)=-\int 
\frac{d^3k}{(2\pi)^3}
V_A({\bf p}-{\bf k})\,\left[\Psi({\bf k},p_0)-\Psi({\bf p},p_0)\right]
+C\, \Psi({\bf p},p_0)\, ,
\label{Schroedinger}
\end{equation}
where $m_R$ is the reduced mass, $E$ is the binding energy, and $p_0$ 
is an {\it
eigenvalue\/} given by
\begin{equation}
p_0^2= 2m_R E \, .
\end{equation}
The constant potential is used to adjust the energy scale.

While Eq.~(\ref{Schroedinger}) was derived for the linear potential 
with the specific
choice of $V_A$ given in Eq.~(\ref{nrpotential2}), it is instructive 
to consider it in
its most general form where $V_A$ is an arbitrary function.  From 
this point of view,
the role of the second term in square brackets in Eq.~(\ref{Schroedinger})
(which arises from the subtraction term), is to insure that the coordinate
space potential $\tilde V_A(r)$ is redefined so that it is zero at the origin;
ie.~Eq.~(\ref{Schroedinger}) is a standard Schr\"odinger equation for 
the potential
\be
\tilde V_L(r)=\tilde V_A(r)-\tilde V_A(0)\, .
\ee
Looking at it this way, we see that {\it any potential $\tilde V_A(r)$ for
which $\tilde V_A(r_o)-\tilde V_A(0)=\infty$, for some $r_o$, gives a confined
system when used with Eq.~(\ref{Schroedinger})\/}.  For example, even 
the choice of
a pure Coulomb-type interaction for $\tilde V_A$,
\begin{equation}
\tilde V_A(r)=-\frac{1}{r}\, ,
\end{equation}
would give confinement.  The subtraction term forces the interaction 
to vanish at
the origin, which requires an infinite shift in the energy (just as in the case
of the linear interaction) forcing the  interaction to go to infinity at
large distances. {\em The role of the subtraction is an essential part of
introducing confinement\/}. This trivial point is worth emphasizing 
because when
we arrive at the relativistic equation, the subtraction term will prove to be
just as crucial as it was in the nonrelativistic Schr\"odinger equation.

We know that Eq.~(\ref{Schroedinger}) confines the
quarks because it was derived from a coordinate space equation which
confines, but it is instructive to see in a simple, direct way how confinement
can be demonstrated directly from the momentum space equation.  To 
see this, first
consider the case when $C=0$, let $\epsilon$ be small but nonzero, 
and rewrite the
Schr\"odinger equation
\begin{equation}
\left[ \frac{{\bf p}^2}{2m_R}-E-\tilde V_A(0)\right]\Psi_A({\bf p},p_0)=-\int
\frac{d^3k}{(2\pi)^3} V_A({\bf p}-{\bf k})\,\Psi_A({\bf k},p_0)\, ,
\label{Schroedinger2}
\end{equation}
where, for the linear potential introduced above,
\be
\tilde V_A(0)=-\frac{\sigma}{\epsilon} \, .
\ee
[For simplicity, we will sometimes refer below to 
Eq.~(\ref{Schroedinger2}) as the {\it
bound state\/} form of the equation.] In coordinate space, the 
potential $\tilde V_A(r)$
approaches zero at large $r$, as illustrated in Fig.~\ref{linear.fig}.  Hence
scattering will take place only if the l.h.s. of this equation has a 
non-trivial
solution, which requires
\begin{equation}
p^2\ge p_0^2 +2m_R \tilde V_A(0)\equiv p_\epsilon^2  \, .
\end{equation}
Note that this implies that
\be
E=\frac{p_0^2}{2m_R}\ge \frac{\sigma}{\epsilon} \to \infty \, ,
\ee
as $\epsilon\to0$, showing that no scattering can take place for finite
energies.  At energies below $1/\epsilon$, only bound states can 
occur.  This is the
demonstration we seek.

Even though Eq.~(\ref{Schroedinger2}) shows that there is no scattering when
$\epsilon\to0$, it is still instructive to write a scattering 
equation for finite
$\epsilon$.  To this end it is convenient to replace $\tilde{V}_L(r)$ by
its counterpart, $\tilde{V}_S(r)$ defined in Eq.~(\ref{2spots}a). 
This potential has
no subtraction, so its momentum space Schr\"odinger equation is simply
\begin{equation}
\left[ \frac{{\bf p}^2}{2m_R}-E\right]\Psi_S({\bf p},p_0)=-\int
\frac{d^3k}{(2\pi)^3} V_S({\bf p}-{\bf k})\,\Psi_S({\bf k},p_0)\, .
\label{Schroedinger2a}
\end{equation}
This will be referred to as the {\it scattering\/} form of the equation.

\begin{figure} 
\centerline{
    \epsfysize=3.0in
\epsffile{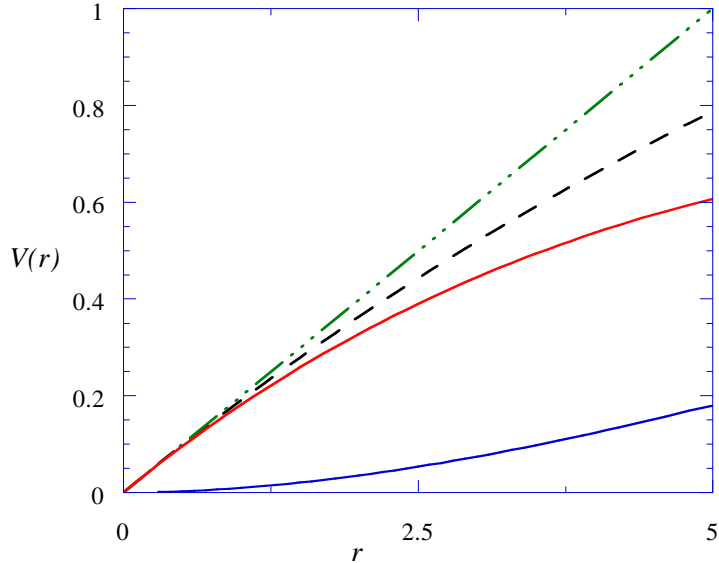}}
\caption{The potentials $\tilde V_S(r)$ (upper solid line), $\tilde 
V_L(r)$ (dashed
line), $\sigma r$ (dot-dashed line), and $\tilde V_0(r)$ (lower solid line) in
coordinate space for $\epsilon=0.1$ and $\sigma=0.2$.}
\label{linear3.fig}
\end{figure}

As stated above, we will assume that the two equations 
(\ref{Schroedinger2}) and
(\ref{Schroedinger2a}) give equivalent results when
$\epsilon$ is very small.  Their equivalence is clear on physical 
grounds, since there
is very little difference, on a sub-atomic scale,  between a barrier 
which is a mile
thick and one which is infinitely thick.  To emphasize this point,
Fig.~\ref{linear3.fig} compares the short distance behavior of the 
potentials $\tilde
V_S$, $\tilde V_L$, and
\bea
\tilde V_0({r})=\tilde V_L({r})-\tilde V_S({r}) \, .
\label{V0}
\eea
As $\epsilon\to0$ for a fixed range of $r$, $\tilde V_0\to0$ and
$\tilde V_S\to\tilde V_L$.   However, a careful mathematical 
treatment of how these two
equations approach the limit as $\epsilon\to0$ presents some subtle issues
\cite{JM92,O97}  which we defer to a subsequent paper.  Our 
arguments in the remainder
of this section are based on simple physical considerations.

\begin{figure} 
\centerline{
    \epsfysize=3.0in
\epsffile{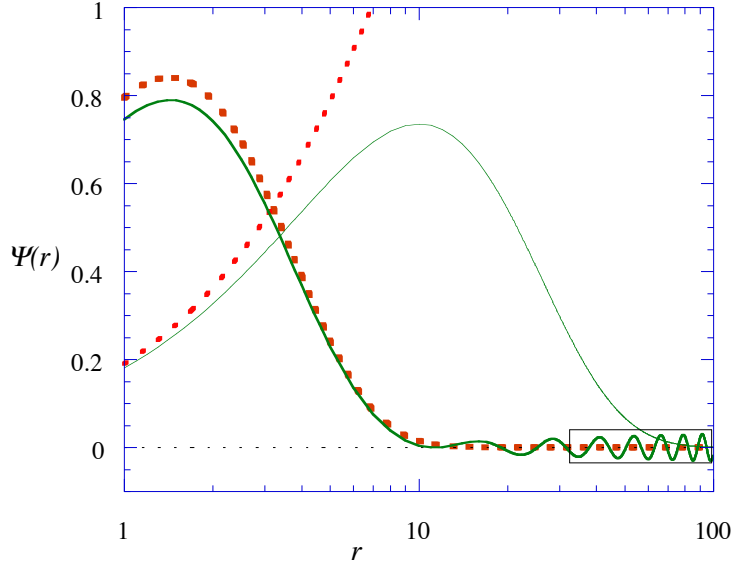}}
\caption{Comparison of possible wave functions $\Psi_A(r)$ (falling 
dotted line) and
$\Psi_S(r)$ (heavy solid line).  [For reference, the potentials 
$\tilde V_S(r)$ (thin solid
line) and $\tilde V_L(r)$ (rising dotted line) are also shown.] The
normalization is chosen so that $\Psi_A\leftrightarrow\Psi_S$, making 
the plane wave tail of
$\Psi_S$ (shown in the box) small. In this example $\eta\simeq0.05$.}
\label{wavecompare.fig}
\end{figure}

In connection with the scattering form (\ref{Schroedinger2a}) we 
introduce a scattering
state wave function defined by
\bea
\Psi_S({\bf p},p_0)=(2\pi)^3\,\eta\;\delta^3(p-p') -
\frac{2m_R\,{M}_S({\bf p},{\bf p}')} {{\bf p}^2-p_0^2}  \, ,
\label{scatwf}
\eea
where ${M}_S$ is the half off-shell scattering amplitude, and ${\bf 
p}'^2=p_0^2$.  The
wave function (\ref{scatwf}) has the form of the usual scattering 
wave function, with the
$\delta$ function describing the asymptotic plane-wave part.  We have 
chosen to multiply
this plane-wave part by a (small) parameter $\eta$.  This parameter 
can be removed by
dividing the wave function and the half off-shell scattering 
amplitude by $\eta$, so
it is, strictly speaking, an arbitrary scale factor.  However, if we 
wish to compare the
scattering solutions to (\ref{Schroedinger2a}) with the bound state 
solutions to
(\ref{Schroedinger2}), it is necessary to choose $\eta$ so that the wave functions are
comparable at small $r$, as illustrated in Fig.~\ref 
{wavecompare.fig}, and this will
require that $\eta$ be very small.  Such a comparison is only 
possible at certain
energies (close to the bound state energies) where the scattering 
solutions are resonant
and therefore much larger at small $r$ than at large $r$.  In 
general, at other energies,
Eq.~(\ref{Schroedinger2a}) will have nonresonant solutions that 
can not be large at small
$r$. Only the resonant solutions of the scattering form 
(\ref{Schroedinger2a}) will converge
to the bound state solutions to Eq.~(\ref{Schroedinger2}), and for these
$\eta$ is very small.  The non-resonant solutions to the scattering 
form are confined to
the large $r$ region, and move off to infinity as
$\epsilon\to0$.  This complicated limiting process will be summarized 
by the equation
\bea
\Psi_S({\bf p},p_0) \leftrightarrow \Psi_A({\bf p},p_0)  \, ,
\label{compare}
\eea
where the $\leftrightarrow$ symbol means that the spectrum of 
resonance scattering
states obtained from (\ref{Schroedinger2a}) converge to the bound 
states obtained from
(\ref{Schroedinger2}), and the nonresonant solutions to (\ref{Schroedinger2a})
can be ignored because they contribute only at infinite energy.

With this insight, we substitute the
scattering wave function (\ref{scatwf}) into the Schr\"odinger equation
(\ref{Schroedinger2}), giving
\bea
{M}_S({\bf p},{\bf p}')&&=\eta\,V_S({\bf p}-{\bf p}')
-2m_R\int
\frac{d^3k}{(2\pi)^3} V_S({\bf p}-{\bf k})\,\frac{{M}_S({\bf
k},{\bf p}')} {{\bf k}^2-p_0^2} \, .
\label{Schroedingerscat}
\eea
Alternatively, we may work directly with the bound state form 
(\ref{Schroedinger2}) of
the equation.  In this case we make the replacement
\bea
\Psi_A({\bf p},p_0)= - \frac{2m_R\,M_A({\bf p},{\bf p}')} {{\bf p}^2-p_0^2}
\label{boundwf}
\eea
and, substituting this into Eq.~(\ref{Schroedinger2}), obtain the 
following equation
\bea
{M}_A({\bf p},{\bf p}')= -2m_R\int
\frac{d^3k}{(2\pi)^3} V_A({\bf p}-{\bf k})\,\left[\frac{{M}_A({\bf
k},{\bf p}')} {{\bf k}^2-p_0^2}-\frac{{M}_A({\bf
p},{\bf p}')} {{\bf p}^2-p_0^2}\right] \, .
\label{Schroedingerbound}
\eea
Following the argument developed above, in the limit $\epsilon\to0$ 
(and $\eta\to0$)
the two amplitudes $M_S$ and $M_A$ should be equivalent.  In the notation of
Eq.~(\ref{compare})
\bea
M_S({\bf p},{\bf p}') \leftrightarrow M_A({\bf p},{\bf p}')  \, .
\label{compare2}
\eea
We will find it convenient to use $M_S$ when $\epsilon$ is very small 
but nonzero, and
to use $M_A$ when we want exact confinement ($\epsilon=0$).  Only
Eq.~(\ref{Schroedingerbound}) has a well defined mathematical limit when
$\epsilon\to0$.   In our subsequent development we will assume that 
either $M_S$ or
$M_A$ may be used with equivalent results.

When $\epsilon=0$, the inhomogeneous term vanishes and there
exist bound states only.  We introduce the {\it vertex\/} function 
$\gamma$ defined by
\be
\Psi_A({\bf p},p_0)= -\frac{2m_R\,{\gamma}({\bf p},p_0)} {{\bf p}^2-p_0^2} \, .
\label{bwfs}
\ee
The Schr\"odinger equation for the vertex function, restoring the constant
interaction term, is
\bea
\gamma({\bf p},p_0)= -2m_R\int \frac{d^3k}{(2\pi)^3}
V_A({\bf p}-{\bf k})\,\left[\frac{\gamma({\bf k},p_0)} {{\bf k}^2-p_0^2}
-\frac{\gamma({\bf p},p_0)}{{\bf p}^2-p_0^2}\right]  +\frac{2m_RC\,\gamma({\bf
p},p_0)}{{\bf p}^2-p_0^2} \, .
\label{Schroedinger3}
\eea

Next, look at this equation when ${\bf p}^2\to p_0^2$.  To this end first write
\be
{\gamma}({\bf p},p_0)={\gamma}(p_0,p_0) +({\bf p}^2-p_0^2)\,{\cal R}
({\bf p},p_0)\, , \label{expan}
\ee
and then substitute this into Eq.~(\ref{Schroedinger3}) [with $C=0$ for the
moment], giving
\bea
{\gamma}({\bf p},p_0)=&&-2m_R\,{\gamma}(p_0,p_0)\int
\frac{d^3k}{(2\pi)^3} V_A({\bf p}-{\bf k})\,\left[\frac{1} {{\bf
k}^2-p_0^2} -\frac{1}{{\bf p}^2-p_0^2}\right]\nonumber\\
&&-2m_R\int
\frac{d^3k}{(2\pi)^3} V_A({\bf p}-{\bf k})\,\left[{\cal R}({\bf k},p_0)
-{\cal R}({\bf p},p_0)\right]
\, .
\label{Schroedingerscat2}
\eea
All terms on the r.h.s. of this equation should be regular as  ${\bf p}^2\to
p_0^2$.  Because of the subtraction, the term involving ${\cal R}$ is finite,
and, because of our choice of $p_0$, only {\it one\/} of the two 
remaining terms
is zero if  $\epsilon$ is {\it finite\/}
\bea
\lim_{{\bf p}^2\to p_0^2}\int\frac{d^3k}{(2\pi)^3} \frac{V_A({\bf 
p}-{\bf k})}{{\bf
k}^2-p_0^2} =&&
-\frac{\sigma}{\epsilon^2}\left(\frac{\epsilon \pm 2i p_0}
{4p_0^2+\epsilon^2}\right) \to {\rm finite} \nonumber\\
\lim_{{\bf p}^2\to p_0^2}\int\frac{d^3k}{(2\pi)^3} \frac{V_A({\bf 
p}-{\bf k})}{{\bf
p}^2-p_0^2} =&& -\frac{\sigma}{\epsilon}\,\lim_{{\bf p}^2\to p_0^2}
\left(\frac{1}{{\bf p}^2-p_0^2}\right)\to \infty\, .
\eea
Hence the subtraction term will be singular unless
\be
{\gamma}(p_0,p_0)=0 \, .  \label{vanish}
\ee
This condition also insures that the constant term is not singular.
We will discuss the physical interpretation of this result in the next section.


\section{Confinement in the spectator formalism}
\subsection{Introduction}

At this point it is very tempting to generalize the nonrelativistic
linear potential Eq.~(\ref{nrpotential}) by simply replacing the three vector
${\bf q}$ by a four vector $q$
\begin{equation}
{\cal V}({\bf q})\stack{?}{\rightarrow} \lim_{\epsilon\rightarrow 0}
\Bigl[V_A(q)-\delta^4(q)\int d^4q'V_A(q')\Bigr]+(2\pi)^3\delta^4(q) C.
\end{equation}
This, seemingly obvious, generalization will not reduce to the correct
nonrelativistic limit because of the unconstrained behavior of the $\int
dq_0'\,V_A(q')$ integral. Lacking a four dimensional expression for the
linear interaction that reduces to the correct nonrelativistic limit, 
we rephrase
our question: Can one find a covariant equation that reduces to the 
Schr\"odinger
Eq.~(\ref{Schroedinger3}) {\em with a linear interaction}?  The confining
relativistic bound state equation should be a relativistic generalization of
Eq.~(\ref{Schroedinger3}).

\begin{figure}
\centerline{
    \epsfysize=1.5in
\epsffile{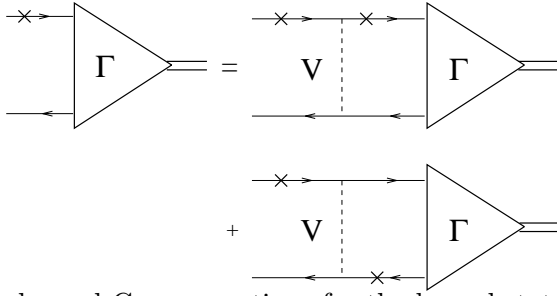}}
\caption{One of the two-channel Gross equations for the bound state
vertex function $\Gamma$.  In this figure the $\times$ means that the 
particle is on
the mass-shell.}
\label{gross.fig}
\end{figure}

A covariant equation with the correct nonrelativistic limit is the Gross
equation\cite{grossref,GROSS2}.  If the two quarks have unequal masses
$m_1>m_2$,  the {\it one channel\/} equation may be used. It has the 
feature that the
four  dimensional loop integrals are constrained so that the heavier 
constituent (with
mass $m_1$ in this example) is restricted to its  positive-energy mass shell
(provided $M_B>0$; see Ref.~\cite{cc}). However, if the particles 
have equal mass
($m_1=m_2=m$) and the mass $M_B$ of the bound state is comparable to 
$m$, a {\it
symmetrized two channel\/} equation should  be used.  This is illustrated in
Fig~\ref{gross.fig}.  In this case an average of the contributions in 
which either
particle 1 (channel 1) or particle 2 (channel 2) are on their positive-energy
mass-shell are included, and this leads to a set of equations in which the two
channels are coupled.  The symmetrized two-channel equation has been used
previously to describe of low energy
$NN$ scattering\cite{GVOH}.  Finally, if the masses are identical and the bound
state mass is very small (ie. $M_B<<m$), as in the chiral limit, then 
a {\it four
channel\/} equation is needed.  The four channel equation is a 
symmetrized version
of the {\it unsymmetrized two channel\/} equation used in 
Ref.\cite{GROSS}.  One of
the purposes of this paper is to improve on this previous work.

\subsection{One channel scattering equations for scalar quarks}

We will begin with the one channel equation.  The momentum and mass
of the quark are $p_1$ and $m_1$, the momentum and mass of the antiquark are
$p_2$ and $m_2$, the total momentum is $P$, and the relative momentum is $p$,
where
\bea
P=&&p_1+p_2=\{M_B,{\bf 0}\}\nonumber\\
p=&&{\frac{1}{2}} \left(p_1-p_2\right) \, . \label{momdef}
\eea
The quark will be on mass-shell, and the symbol ${p}_1^+$ will be 
used to denote
the particle on its {\it positive\/} energy mass-shell, (ie. 
${p}_1^{+\,2}=m_1^2$ and
$({p_{1}^+})_0=E_1(p)=\sqrt{m_1^2+{\bf p}^2}$).  The scattering 
amplitude ${\cal
M}({p}^+_1,p_2,{p}'^+_1,p'_2)$ is denoted ${\cal M}_{11}({\bf p},{\bf 
p}',P)$, or
in the one channel case where there can be no confusion, simply by 
${\cal M}({\bf
p},{\bf p}',P)$.   Then, introducing a relativistic generalization of 
the potential
$V_S$, the one channel equation for the scattering of scalar 
``quarks'' ($m_1>m_2$) can
be written
\bea
{\cal M}_S({\bf p},{\bf p}',P)=&&\eta V_S({\bf p},{\bf p}',P) -
\frac{2m_1m_2}{(2\pi)^3}\int\frac{d^3k }{E_1(k)}\;\frac{V_S({\bf 
p},{\bf k},P)\,{\cal
M}_S({\bf k},{\bf p}',P)}{m_2^2-(P-{k}^+_1)^2} \nonumber\\
&&+\frac{ 2m_2C\,{\cal M}_S({\bf p},{\bf p}',p)}{m_2^2-(P-{p}^+_1)^2} \, .
\label{Gross01}
\eea
This equation is the relativistic generalization of 
Eq.~(\ref{Schroedingerscat}).

Alternatively, the bound state form of the scattering equation is
\bea
{\cal M}_A({\bf p},{\bf p}',P)= && -\frac{2m_1m_2}{(2\pi)^3} 
\int\frac{d^3k}{E_1(k)}
V_A({\bf p},{\bf k},P)\;\left[\frac{{\cal M}_A({\bf k},{\bf
p}',P)}{m_2^2-(P-{k}^+_1)^2}-\frac{{\cal M}_A({\bf p},{\bf
p}',P)}{m_2^2-(P-{p}^+_1)^2}\right]
\nonumber\\ &&+\frac{ 2m_2C\,{\cal M}_A({\bf p},{\bf
p}',P)}{m_2^2-(P-{p}^+_1)^2}\, .
\label{Gross0}
\eea
This is the analogue of Eq.~(\ref{Schroedingerbound}) and has a smooth limit as
$\epsilon\to0$.  The kernels $V_S$ and $V_A$ will be specified later (see
Eqs.~(\ref{VS}) and (\ref{nqtothe4}) below).    Equations (\ref{Gross01}) and
(\ref{Gross0}) will be our starting points for this section.

\subsection{One channel bound state equation for scalar quarks}

In the vicinity of a bound state of mass $M_B$, or a very narrow 
resonance with mass
and width $M_B=M_R+iM_I$,  the scattering amplitude has the form
\be
{\cal M}_X({\bf p},{\bf p}',P)= -\frac{\Gamma_X({\bf p},M_B)\,\Gamma_X({\bf
p}',M_B)}{M_B^2-P^2} + {\cal R}_X({\bf p},M_B)\, , \label{gform}
\ee
where $X=A$ or $S$, depending which of the two forms (\ref{Gross01}) 
or (\ref{Gross0}) we
are using.  If $\epsilon$ is finite and we are using 
Eq.~(\ref{Gross01}), the width
$M_I\ne0$.  If we use Eq.~(\ref{Gross01}) the width is zero for all 
states with mass
below some critical mass $M_\epsilon \to\infty$ as $\epsilon\to0$.

Substituting the form (\ref{gform}) into either Eq.~(\ref{Gross01}) or
Eq.~(\ref{Gross0}), and equating residues at the pole (real or 
complex) gives the bound
state equations for the vertex functions $\Gamma_X$:
\bea
\Gamma_S({\bf p},M_B)=-2m_1m_2&&\int \frac{d^3k\;V_S({p},{k},M_B)}{(2\pi)^3
\,E_1(k)}\frac{\Gamma_S({\bf k},M_B)}{m_2^2-(M_B-{k}^+_1)^2}\nonumber\\
&&+\frac{ 2m_2\,C\,\Gamma_S({\bf p},M_B)}{m_2^2-(M_B-{p}^+_1)^2}\, , 
\label{Gross0a}\\
\Gamma_A({\bf p},M_B)=-2m_1m_2&&\int \frac{d^3k\;V_A({p},{k},M_B)}{(2\pi)^3
\,E_1(k)}
\left[\frac{\Gamma_A({\bf k},M_B)}{m_2^2-(M_B-{k}^+_1)^2}
-\frac{\Gamma_A({\bf p},M_B)}{m_2^2-(M_B-{p}^+_1)^2}\right]\nonumber\\
&&+\frac{ 2m_2\,C\,\Gamma_A({\bf p},M_B)}{m_2^2-(M_B-{p}^+_1)^2}\, . 
\label{Gross0b}
\eea
where we use a mixed notation with $M_B$ denoting both the mass and the four
vector $\{M_B,0\}$, the difference being clear from the context.

As with the scattering amplitudes, the two vertex functions are 
equivalent in the limit
$\epsilon\to0$
\bea
\Gamma_S({\bf p},M_B) \leftrightarrow \Gamma_A({\bf p},M_B)  \, ,
\label{compare3}
\eea
but the vertex function $\Gamma_A$ is more convenient to calculate in the limit
$\epsilon\to0$.

\subsection{Normalization condition}

The bound state equation and the normalization condition for the 
bound state wave
function can be derived from a nonlinear form of 
Eq.~(\ref{Gross01})\cite {AGVO}.
In this paper the derivative $\partial V_S/\partial P_\mu=0$ in the 
rest frame, so the
result is
\be
2P^\mu=\frac{\partial}{\partial P_\mu}\int \frac{d^3 k}{(2\pi)^3\,E_1(k)}
\left\{\frac{\Gamma_S({\bf k},M_B)\,\Gamma_S({\bf
k},M_B)}{m_2^2-(P-{k}^+_1)^2}\right\}
\, . \label{norm1}
\ee
In view of the relation (\ref{compare3}) this relation can also be written
\bea
2P^\mu=&&\frac{\partial}{\partial P_\mu}\int \frac{d^3 k}{(2\pi)^3\,E_1(k)}
\left\{\frac{{\Gamma}_A({\bf k},M_B)\,{\Gamma}_A({\bf 
k},M_B)}{m_2^2-(P-{k}^+_1)^2}
\right\}\nonumber\\
=&&\int \frac{d^3 k}{(2\pi)^3\,E_1(k)}
\frac{{\Gamma}_A({\bf k},M_B)\,2(P-{k}^+_1)^\mu\,{\Gamma}_A({\bf k},M_B)}
{\left(m_2^2-(P-{k}^+_1)^2\right)^2} \, .
\eea
This is a familiar result, which will be generalized to the spin 1/2 
case later.

\subsection{Symmetrized two channel equation for equal mass scalar quarks}

If the quarks have equal mass ($m_1=m_2=m$), and the bound state mass 
is positive and
not too small, a symmetrized two channel equation is needed. The two 
channels will be
labeled 1 and 2 depending on whether the quark or antiquark is on 
mass-shell, and the
symbol
${p}_1^+$ denotes that the particle is on its {\it positive\/} energy 
mass-shell, (ie.
${p}_1^{+\,2}=m^2$ and ${p_0^+}=E(p)=\sqrt{m^2+{\bf p}^2}$).  Starting from
Eq.(\ref{Gross0b}), and suppressing the subscript $A$, the vertex 
functions for the two
channels are denoted
\begin{equation}
\Gamma_1({\bf p},M_B)=\Gamma({p}^+_1,p_2),\qquad
\Gamma_2({\bf p},M_B)\equiv\Gamma(p_1,{p}^+_2)\, .
\end{equation}
With this notation, the symmetrized two channel equation for equal mass scalar
``quarks'' with a confining interaction can be written
\bea
\Gamma_i({\bf p},M_B)=-m^2&&\sum_j\int \frac{d^3k\;V_{ij}({p},{k})}{(2\pi)^3
\,E_j(k)}
\left[\frac{\Gamma_j({\bf k},M_B)}{m^2-(P-{k}^+_j)^2}
-\frac{\Gamma_i({\bf p},M_B)}{m^2-(P-{p}^+_i)^2}\right]\nonumber\\
&&+\frac{ 2mC\,\Gamma_i({\bf p},M_B)}{m^2-(P-{p}^+_i)^2}\, ,
\label{Gross2}
\eea
where $i$ and $j$ label which of the two quarks is on-shell, and
\be
{k}^+_j=\{E(k),(-)^{j+1}\,{\bf k}\}\label{momentum1}
\ee
is the momentum of the on-shell quark.   Note that the strength of 
the $V_{ij}$ term has
been multiplied by 1/2, reflecting the fact that the interaction is 
an {\it average\/} of
the strengths in two channels which are equal in the nonrelativistic 
limit. This
equation uses the same subtraction for both the $i=j$ and the
$i\ne j$ terms.  This prescription differs from that previously used 
in Ref.\cite{GROSS}.
In this work the kernel below will not, in general, be singular when 
$i\ne j$, and the
subtraction used above is sufficient to preserve the nonrelativistic 
limit (see below).

In order to complete the description we need to specify the form of
covariant interaction $V_{ij}$. A natural choice that reduces to the
correct nonrelativistic limit is \cite{GROSS}
\begin{equation}
V_{ij}({p},{k})\equiv V_A(q_{ij})=
-\frac{8\pi\sigma}{(q_{ij}^2-\epsilon^2)^2}\, ,
\label{V1}
\end{equation}
where the four-momentum transfer depends on whether or not $i=j$:
\bea
q_{11}^2=&&q_{22}^2=\left(E(k) -E(p)\right)^2-({\bf k}-{\bf p})^2 \nonumber\\
q_{12}^2=&&q_{21}^2=\left(M_B-E(k) -E(p)\right)^2-({\bf k}+{\bf p})^2 \, .
\eea
A similar form could be used for the kernel $V_S$ (which we will not need)
\begin{equation}
V_{ij}({p},{k})\equiv V_S(q_{ij})=
-8\pi\sigma\left\{\frac{1}{(q_{ij}^2-\epsilon^2)^2} +
\frac{4\epsilon^2}{(q_{ij}^2-\epsilon^2)^3}\right\}\, .
\label{VS}
\end{equation}
However, the form (\ref{V1}) has two drawbacks. First, at large ${\bf 
p}\simeq{\bf k}$
the kernel converges slowly, and the equation is ultraviolet divergent.  In
Ref.~\cite{GROSS} a form factor was introduced to regularize this divergence.
Second, using this form it is difficult to regularize the infrared ($q^2=0$)
singularities that appear in the $\epsilon=0$ limit. In the 
nonrelativistic case the
infrared singularity occurs only at ${\bf q}=0$ and can be regulated 
by  the $\delta$
function subtraction in Eq.~(\ref{nrpotential}). However, in the 
relativistic case
infrared singularities occur not only when $q^{\mu}=0$, but also (for 
the $i\ne j$
kernels) when the momentum transfer is light-like, so that $q^2=0$ 
but $q^{\mu}\neq 0$.
These ``{\em off-diagonal\/}'' singularities are not  regulated by 
the subtraction term,
and their removal spoils the simplicity of this approach\cite{GROSS}.

Since the role of $V_A$ is to model the linear interaction, and the 
principle requirement
is that it reduce to the correct nonrelativistic limit, both of these
problems are eliminated very simply if $V_A$ is defined as follows
\begin{equation}
V_A(q_{ij})= -\frac{8\pi \sigma}{q_{ij}^4+(P\cdot q_{ij})^4/P^4}
\label{nqtothe4}
\end{equation}
where $P$ is the total four-momentum of the bound state. This form
has the following advantages:

\vspace*{0.1in}

\hang{(i)} the denominator is not singular unless both $q^2$ and $P\cdot q$ are
zero, so the singularities are restricted to $q^{\mu}=0$;

{(ii)} no ultraviolet regularization is needed;

\hang{(iii)} the interaction does {\em not} depend on the bound state 
momentum $P$ in
the bound state rest frame; and

  {(iv)} it has the  correct nonrelativistic dependence on ${\bf q}^2$.

\vspace*{0.1in}

\noindent One disadvantage of the form (\ref{nqtothe4}) is its 
dependence on the total
momentum $P$ of the particle pair.  However, since since this kernel confines
particles in pairs that can not be separated, they are naturally 
associated as a pair
and we do not view this as a serious limitation.  Another feature of the form
(\ref{nqtothe4}) is that its off-diagonal couplings are singular only when
$W=2E(p)$ (because ${\bf k} + {\bf p}=0$ also).  This is only possible 
for excited
states and, as we will prove below, confinement requires the vertex
function to be zero at this point, controlling this singularity 
automatically.

The introduction of the definition (\ref{nqtothe4}) considerably
simplifies the solution of the relativistic equations (\ref{Gross2}), but will
introduce electromagnetic interaction currents if the photon 
four-momentum is not zero.
These will be discussed in a subsequent paper.

Both Eqs.~(\ref{Gross0b}) and (\ref{Gross2}) have the correct nonrelativistic
limit with confinement.  Consider the one-channel Eq.~(\ref{Gross0b}) 
first, and let
$m_1$ and $m_2\to\infty$.  Then the energy transferred by the on-shell quark,
$E_1(k)-E_1(p) \to0$ and $V_A(q_{11})\to V_A({\bf q})$.  Furthermore, if
$M_B=m_2+m_1+E$, then to first order in the small quantities ${\bf 
k}^2$ and $m_RE$,
the relativistic propagator reduces to
\be
\frac{1}{m_2^2-k_2^2}\to \frac{m_R}{m_2({\bf k}^2-2m_RE)}\, ,
\ee
and substituting this into Eq.~(\ref{Gross0b}) gives
Eq.~(\ref{Schroedinger3}).  In the two channel case $q_{11}\to 
q_{12}$ as $m\to\infty$ and
the kernels $V_{11}\to V_{12}$.  Since the subtraction in the two 
channels is also
identical, the contributions from the two channels are equal and the 
coupled equations
reduce to the single Eq.~(\ref{Schroedinger3}).

\subsection{Proof of confinement}

While one can visualize the potential in the nonrelativistic case and 
get a picture of
the physics, it is less possible to visualize the covariant 
interaction.  {\em What are
the criteria with which one can judge whether a given interaction 
really confines\/}?
If the particles are bound in a state of total mass  larger than the 
sum of the masses
of the constituents ($M_B>m_1+m_2$), the  bound state could in 
principle decay into free
constituents.  Confinement prevents this from happening in one of two 
possible ways:
{(i)} the quark propagators will not have any physical mass poles 
\cite{CETIN}, or,
as we will now prove for this model, {(ii)} the vertex function will 
vanish when the
quarks are simultaneously on-shell.

\begin{figure}
\centerline{
    \epsfxsize=2in
\epsffile{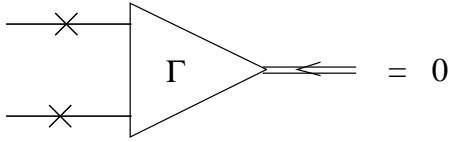}}
\caption{The confinement condition for the Gross vertex function.}
\label{conf.fig}
\end{figure}

The proof is identical to the nonrelativistic proof given above and we will
summarize it only for the one channel equation.  Setting $C=0$, the one channel
bound state Eq.~(\ref{Gross0b}) can be written
\bea
\Gamma_A({\bf p},M_B)=&&-2m_1m_2\int \frac{d^3k\;V_A(p,k)}{(2\pi)^3 \,E_1(k)}
\left[\frac{\Gamma_A({\bf k},M_B)-\Gamma_A({\bf 
p},M_B)}{m_2^2-k_2^2}\right]\nonumber\\
&&+2m_1m_2\Gamma_A({\bf p},M_B)\int \frac{d^3k\;V_A(p,k)}{(2\pi)^3
\,E_1(k)}\left\{\frac{p_2^2-k_2^2}{\left(m_2^2-p_2^2\right)
\left(m_2^2-k_2^2\right)}\right\}\, .
\label{Gross3}
\eea
Since the first quark is on-shell, the second quark is on its {\it 
positive\/} energy
mass shell when the magnitude of the relative three-momentum $|{\bf 
p}|={p}_0$ is
\begin{equation}
\sqrt{m_1^2+{p}_0^2}+\sqrt{m_2^2+{p}_0^2}=M_B\, .
\end{equation}
This occurs when ${p}_0^2$ is given by
\begin{equation}
4M_B^2\,{p}_0^2=\left[M_B^2-(m_1+m_2)^2\right]\left[M_B^2-(m_1-m_2)^2\right]\, .
\end{equation}
As in the nonrelativistic case, the singularity at ${\bf p}={\bf k}$ 
is integrable,
and hence the second term on r.h.s. of Eq.~(\ref{Gross3}) will be 
singular at ${\bf
p}={p}_0\,{\bf \hat p}\equiv {\bf p}_0$ (where ${\bf \hat p}$ is a 
unit vector in the
direction of ${\bf p}$) unless
\begin{equation}
\Gamma({\bf p}_0,M_B)=0\, . \label{confcond}
\end{equation}
Therefore, {\it the vertex function vanishes  when both particles are 
on their mass
shell\/}.  This condition is illustrated diagrammatically in 
Fig.~\ref{conf.fig}.

Note that the subtraction term in Eqs.~(\ref{Gross2}) and 
(\ref{Gross3}) plays two
central roles: (i) it regularizes the singular interaction at ${\bf
p}={\bf k}$ and and makes it zero at $r=0$, and (ii) it is singular 
when $p_2^2\to
m_2^2$, forcing condition (\ref{confcond}).  {\it The subtraction 
term is essential to the
self consistent description of confinement.\/}  As in the 
nonrelativistic case the proof
did not depend on the specific form of the interaction.

We now discuss how confinement affects the stability of
bound states under external disturbances.

\subsection{Excitation of bound states}
\label{scattering}

\begin{figure}
\centerline{
    \epsfxsize=2.2in
\epsffile{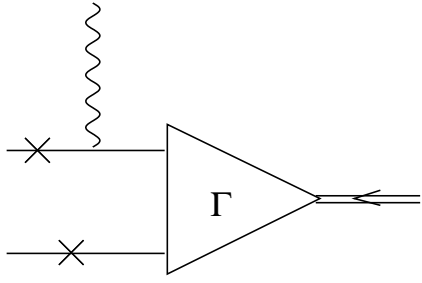}}
\caption{The Born term, which can not exist if the quarks are confined.}
\label{born}
\end{figure}

\begin{figure}
\centerline{
    \epsfxsize=3.5in
\epsffile{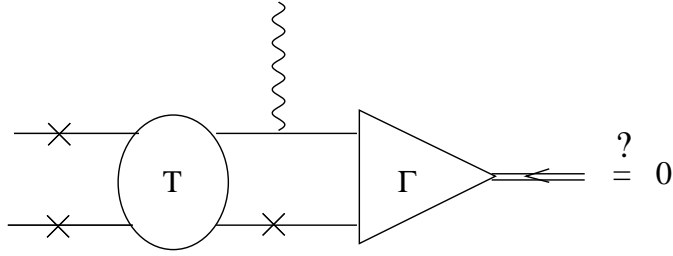}}
\caption{Can an external photon probe disintegrate the bound state?}
\label{conf2.fig}
\end{figure}

A consistent description of confinement implies that two free quarks can not be
liberated from a bound state, even under the influence of an 
energetic external photon or
other probe.  This requirement implies that the usual Born term (shown 
in  Fig.~\ref{born})
is either cancelled by the rescattering term, or is a diagram that 
does not exist in the
formalism.  If the Born term does not exist, the rescattering term, 
illustrated in
Fig~\ref{conf2.fig}, must be zero if the final state  quarks are all 
on-shell.  How are
these restrictions built into the formalism?

When particles are confined there are no free two-particle states and 
the two-body
propagator must always include an infinite number of interactions. 
Since there are no free
particle states, a perturbation theory for confined particles built 
around the free
propagator can not be constructed.  This feature is built-in 
automatically if the two
body propagators satisfy {\it homogeneous\/} integral equations with 
{\it no free
particle contribution\/}.

To illustrate these ideas we review the formalism for the scattering 
amplitude, and its
relation to the two-body propagator.  It is convenient to work with 
the scattering
form of the equation.  In operator notation, Eq.(\ref{Schroedinger2a}) is:
\bea
M({\bf p},{\bf p}',P)=&&\eta\,V(p,p') - V(p,k)G_0({\bf k},{\bf k}',P) 
M({\bf k}',{\bf p}',P)
\nonumber\\
=&& \eta\,V(p,p') -  M({\bf p},{\bf k},P) G_0({\bf k},{\bf k}',P) V(k',p')\, ,
\label{GandM}
\eea
where $G_0({\bf k},{\bf k}',P)$ is the free two body propagator 
[containing a factor of
$\delta^3(k-k')$], integration over $d^3k$ and $d^3k'$ is implied, 
and we have dropped the
subscript $S$ for simplicity.  The parameter $\eta$  was introduced 
in the discussion
following Eq.(\ref{scatwf}) and is very small, approaching zero as 
$\epsilon\to0$.

Now the dressed propagator, $G$ is related to the scattering amplitude $M$ by
\bea
G({\bf p},{\bf p}',P)=&&\zeta\,G_0({\bf p},{\bf p}',P) - G_0({\bf 
p},{\bf k},P) M({\bf
k},{\bf k}',P) G_0({\bf k}',{\bf p}',P) \, ,
\label{GMandG0}
\eea
where $\zeta$, to be determined, is a parameter proportional to the 
strength of the free
particle scattering.  If the potential confines there should be {\it 
no inhomogeneous
term\/} and $\zeta=0$.  To determine $\zeta$ and the equation for 
$G$, substitute
(\ref{GandM}) into (\ref{GMandG0}) giving
\bea
G({\bf p},{\bf p}',P)=&&\zeta\,G_0({\bf p},{\bf p}',P) - \eta 
G_0({\bf p},{\bf k},P) V(k,k')
G_0({\bf k}',{\bf p}',P) \nonumber\\
&& + G_0({\bf p},{\bf k},P) V(k,k')  G_0({\bf k'},{\bf k}'',P) M({\bf
k}'',{\bf k}''',P) G_0({\bf k}''',{\bf p}',P) \nonumber\\
=&&\zeta\,G_0({\bf p},{\bf p}',P) +(\zeta-\eta)\, G_0({\bf p},{\bf 
k},P) V(k,k')
G_0({\bf k}',{\bf p}',P)\nonumber\\
&&-G_0({\bf p},{\bf k},P) V(k,k')  G({\bf k'},{\bf p}',P)\, .
\label{GandM2}
\eea
The second term is eliminated by choosing $\zeta=\eta$, and gives 
familiar equations
for the dressed propagator
\bea
G({\bf p},{\bf p}',P)=&&\eta\,G_0({\bf p},{\bf p}',P) - G_0({\bf 
p},{\bf k},P) V(k,k')
G({\bf k}',{\bf p}',P)
\nonumber\\
=&& \eta\,G_0({\bf p},{\bf p}',P) - G({\bf p},{\bf k},P) V(k,k')
G_0({\bf k}',{\bf p}',P)\, ,
\label{GandM3}
\eea
where the second form parallels the second form of Eq.~(\ref{GandM}).

The interpretation of equations (\ref{GMandG0}) and (\ref{GandM3}) 
for the dressed
propagator follows from the interpretation of Eq.~(\ref{GandM}) for 
the scattering
amplitude.  {\it As $\epsilon\to0$, the parameter $\eta\to0$ and the 
inhomogeneous term
vanishes.  In this limit both the scattering amplitude and the 
propagator satisfy
homogeneous equations.\/}

The inelastic scattering amplitude can be obtained from the dressed 
propagator by striping
off the final free propagators, and is \cite{GROSS2}
\bea
{\cal J}({\bf p},P,q)=&& G_0^{-1}({\bf p},{\bf k},P+q) G({\bf k},{\bf p}',P+q)
J(P+q,P)\Psi(P)
\nonumber\\
=&&\Bigl\{\eta+ M({\bf p},{\bf k},P+q)G_0({\bf k},{\bf p}',P+q) 
\Bigr\}\, J(P+q,P)\Psi(P)
\, .  \label{current1}
\eea
Here the first term proportional to $\eta$ is the Born term shown in
Fig.~\ref{born}, and we see that {\it there is no Born term in the 
limit of exact
confinement\/} (ie. $\eta=0$).  Furthermore, in the presence of confinement the
scattering matrix satisfies the same homogeneous equation satisfied by the bound
states [Eq.~(\ref{GandM}) with $\eta=0$], and an extension of the 
proof given in
Subsec.~F above shows that the {\it scattering matrix in Fig.~\ref{conf2.fig}
must be zero if both final state quarks are on shell\/}.

We have constructed a self-consistent description of confinement 
within the context of
relativistic field theory.


\subsection{Generalization to Fermions}
\label{fermions}

If the quarks have spin, the kernel in the spectator equation will be 
an operator in the
Dirac space of the two quarks.  This operator can be written
\begin{eqnarray}
{\cal V}(p,k)=\sum_{i=1}^3 \alpha_i\,O_{i1}\,O_{i2}\;V_i(p,k) \, ,
\label{kernel0}
\end{eqnarray}
where the Dirac matrices $O$, which operate on the Dirac indices of particles
1 and 2, describe the spin dependent structure of quark-antiquark 
interaction.  The
$\alpha_i$ are parameters determined either empirically (by fitting 
the spectrum), from
lattice calculations, or from the theory. In this paper we consider 
only three possible
spin structures: scalar $O_{1j}=\openone_j$, pseudoscalar
$O_{2j}=\gamma_{5j}$, and vector $O_{3j}=\gamma_{\mu j}/{2}$.  With 
this notation
the one channel spectator equation for spin 1/2 particles with constant masses
$m_1>>m_2$ is given by
\begin{equation}
\Gamma(p,P)=-\int \frac{d^3k}{(2\pi)^3E_1(k)}\, \sum_i V_i(p,k)
O_{i1}(m_1+{\s k}^+_1)\left\{\frac{\Gamma(k,P)}{m_2^2-k_2^2}\right\}
(m_2-\s k_2)O_{i2},
\label{Gross1}
\end{equation}
where the quark has mass $m_1$ and is on shell, so that
${k}_1^{+\,2}=m_1^2={p}_1^{+\,2}$, and the antiquark  has mass $m_2$. 
Therefore, the
momentum transfered by the interaction is
\begin{equation}
({p}^+-{k}^+)^2=(E_1(p)-E_1(k))^2-({\bf p}-{\bf k})^2\equiv q\, .
\end{equation}

As in the nonrelativistic case, we consider a kernel composed of 
linear, constant, and
one gluon exchange (OGE) pieces. The interaction kernel for the 
linear part of the
potential, ${\cal V}_{L}$, is
\begin{eqnarray}
{\cal V}_{L}(p,k)&\equiv&\sum_{i=1}^3 \alpha_{Li}\,O_{i1}\,O_{i2}V_L(p,k),
\nonumber\\
&=&\left(\alpha_s
\openone_1\openone_2+\alpha_{ps}\gamma_{51}\gamma_{52}+{\frac{1}{4}}\,
\alpha_v\, \gamma_{\mu1}\gamma^\mu_2\right)\,V_L(p,k)\, ,
\label{kernel1}
\end{eqnarray}
where $V_L(p,k)$ is
\begin{eqnarray}
V_{L}(p,k)= V_A(q_{11}(p,k))-E_1(k)\,\delta^3(p-k)\int 
d^3k'\frac{V_A(q_{11}(p,k'))}
{E_1(k')} \, .
\label{kernel1a}
\end{eqnarray}
In this work we employ a pure scalar linear interaction,
$\alpha_s=1,\alpha_{ps}=\alpha_v=0$, but in later calculations the coefficients
$\alpha_i$ will be determined empirically. The one gluon exchange and constant
interactions will be pure vector
\begin{eqnarray}
{\cal V}_{g}(q)&=& \gamma_{\mu1}\,\gamma_{\nu2}\,V_g^{\mu\nu}(q) \nonumber\\
{\cal V}_{c}(q)&=& \gamma_{\mu1}\,\gamma^{\mu}_2\, C\, ,
\label{kernel2}
\end{eqnarray}
where
\begin{eqnarray}
V_{g}^{\mu\nu}(q)&\equiv&-(g^{\mu\nu}-\frac{q^\mu 
q^\nu}{q^2})\,V_g(q)\nonumber\\
&=&-(g^{\mu\nu}-\frac{q^\mu q^\nu}{q^2})\,\frac{1}
{q^2-\Lambda^2}\,\frac{d\;16\pi^2/3}{\ln(\tau + |q^2|/\lambda_{QCD}^2)}\,  ,
\label{kernel3}
\end{eqnarray}
where $d=12/(33-2N_f)=12/27$, the color factor of 4/3 has been 
included, $\Lambda=1$ GeV,
$\tau$ = 2, and $\lambda_{QCD}$ = 200 MeV. In previous work 
\cite{GROSS} quark propagators
with constant masses were used.  In this work we parametrize the 
quark propagator by
\begin{equation}
S(p)= \frac{1}{m(p)-\not\! p}\, ,
\label{prop}
\end{equation}
where $m(p)$ is a mass function for the quark, to be defined later.

\begin{figure}
\centerline{
    \epsfxsize=3.6in
\epsffile{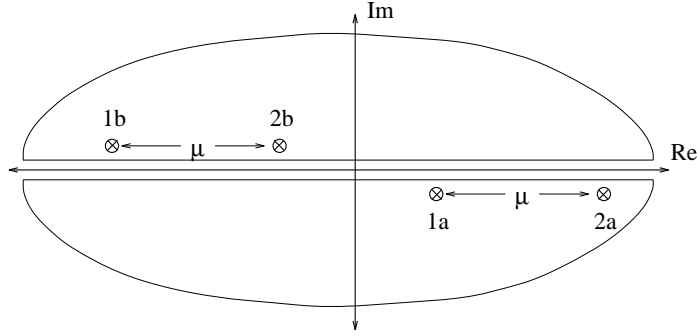}}
\caption{Propagator poles in the complex $k_0$ plane.}
\label{poles}
\end{figure}

If the constituents are identical or close in mass and the equations 
are to be applied
to the description of nearly massless bound states, the {\it four 
channel\/} equation
should be used. Numerical solutions the four channel equation will be 
presented in
this work.

The four channels are defined by the constraints in the four-momenta $k_1$ and
$k_2$ arising from the requirement that {\it both\/} the quark and 
the antiquark be
constrained to {\it both\/} their positive {\it and negative\/} 
energy mass-shells.  A
formal way to obtain the equations is to integrate over the internal 
energy $k_0$ by
averaging the contributions from the quark and antiquark poles in 
{\it both\/} the
upper and lower half $k_0$ complex plane, as illustrated in 
Fig.~\ref{poles}. This
averaging is needed to ensure charge conjugation 
(particle-antiparticle) symmetry, and
leads to four coupled equations.  However, even though the form of 
the equations is
obtained in this way, we emphasize that the equations are theoretically
justified by the argument that the singularities in the interaction 
kernel omitted in
this procedure tend to be cancelled by other higher order terms which 
would otherwise
have been neglected, and that this leads to covariant equations with 
the correct
nonrelativistic limit.  The inclusion of the negative energy poles, 
neglected in
other applications of the symmetrized equations\cite{GVOH}, is 
required in cases
where $P\to0$\cite{GROSS}.

The four constraints are conveniently identified by the notation
\bea
{k}_j^s=\{sE(k), (-)^{j+1}\,{\bf k}\}\, .\label{momentum2}
\eea
which generalizes that introduced in Eq.~(\ref{momentum1}).  Here the 
superscript
$s=\pm$ denotes either the positive or negative energy mass shell constraints.
Then, introducing the projection operators
\begin{equation}
\Lambda(k)= m(k)+\s k \, ,
\label{project}
\end{equation}
and defining the four channel vertex functions
\bea
\Gamma_1^s({\bf p},M_B)=&&\Gamma(p_1^s,p_2) \nonumber\\
\Gamma_2^s({\bf p},M_B)=&&\Gamma(p_1,p_2^s)\, , \label{fourvertex}
\eea
and wave functions
\bea
\Psi_1^s({\bf p},M_B)=&&\frac{\Lambda(p_1^s)\Gamma^{s}_1({\bf 
p},M_B)\Lambda(-p_2)}
{m(p_2)^2-p_{2}^2}   \nonumber\\
\Psi_2^s({\bf p},M_B)=&&\frac{\Lambda(p_1)\Gamma^{s}_2({\bf 
p},M_B)\Lambda(-p_2^s)}
{m(p_1)^2-p_{1}^2}\, ,
\eea
permits us to write the four-channel spectator equation in the 
following compact
form
\begin{eqnarray}
\Gamma_i^{s}({\bf p},M_B)=&-&\frac{1}{2}\sum_{jr}\int 
\frac{d^3k}{(2\pi)^3\,2E(k)}
\Biggl\{V_{ij}^{sr}(p,k)
\left[\Psi_j^r({\bf k},M_B)  - \Psi_i^s({\bf p},M_B) \right]\nonumber\\
&-&2\delta_{ij}\delta_{sr}V_g^{\mu\nu}(p-k)\gamma_\mu \Psi_j^r({\bf k},M_B)
\gamma_\nu\Biggr\}-C\,\gamma_\mu \Psi_i^s({\bf p},M_B)
\gamma^\mu \, ,
\label{4channel}
\end{eqnarray}
where the r.h.s. of the equation now sums over both positive and 
negative energy
contributions ($r=\pm$) from {\it each\/} quark ($j=\pm$).  The Kronecker
$\delta_{ij}\delta_{sr}$ functions restrict the one gluon exchange 
interaction to the
diagonal channels (where the same particle is on the same mass shell 
before and after
the interaction). Inclusion of the one gluon exchange in off-diagonal 
channels leads
to numerical instabilities, which in principle can be handled  by 
using more grid
points in numerical integrations.  Restricting this interaction to 
diagonal channels
eliminates these singularities from the gluon propagator.

\subsection{Charge conjugation invariance}

The final task is to show that Eq.~(\ref{4channel}) is invariant 
under the charge
conjugation operation
\be
\Gamma^{\cal C}(p_1,p_2)={\cal C}\Gamma^{\rm T}(p_2,p_1) {\cal C}^{-1}\, .
\label{gammatrans}
\ee
This is done by proving that both $\Gamma$ and $\Gamma^{\cal C}$ satisfy the
same equation.

First note that, when particle 1 is on shell, interchange of $p_1$ 
and $p_2$ gives
\bea
\Gamma_1^s({\bf p},M_B)=\Gamma(p_1^s,p_2)\to \Gamma(p_2,p_1^s)=\Gamma_2^s(-{\bf
p},M_B) \label{wavetrans}
\eea
and is equivalent to $1\leftrightarrow2$ and ${\bf p}\to -{\bf p}$.  Then
\bea
\Psi_1^{s\,{\cal C}}({\bf p},M_B)=&&{\cal C}\Psi_2^{s\,{\rm T}}(-{\bf 
p},M_B){\cal
C}^{-1}
\nonumber\\
\Psi_2^{s\,{\cal C}}({\bf p},M_B)=&&{\cal C}\Psi_1^{s\,{\rm T}}(-{\bf 
p},M_B) {\cal
C}^{-1}\, .
\eea
Finally, the Dirac direct products $\openone\otimes\openone$,
$\gamma_\mu\otimes\gamma^\mu$, and $\gamma_5\otimes\gamma_5$ are 
invariant under
${\cal C}$.  Hence, changing ${\bf k}\to -{\bf k}$ and performing the 
transformations
(\ref{gammatrans}) and (\ref{wavetrans}), shows that 
Eq.~(\ref{4channel}) is also
invariant.  Therefore the charge conjugation eigenstates, labeled by $\eta=\pm$
\begin{eqnarray}
\Gamma_\eta^{s}({\bf p},M_B)=&&\Gamma_1^s({\bf p},M_B)+\eta\,
\Gamma_2^{s\,{\cal C}}({\bf p},M_B) \, ,
\label{4channelcc}
\end{eqnarray}
are solutions of the equation and charge conjugation symmetry is proved.

\subsection{Dynamical quark mass}

The dynamical quark mass function is the solution of the 
Dyson-Schwinger equation.
In NJL-type models, this {\it one-body\/} equation for the 
spontaneous generation
of quark mass and the {\it two-body\/} bound state equation for a state of zero
mass become identical in the chiral limit (when the bare quark mass 
is zero). In
this limit the quark mass function and the bound state wavefunction 
for a massless
pseudoscalar bound state are identical, and spontaneous symmetry 
breaking assures
the existance of a massless pseudoscalar bound state.

In this paper we adopt a slightly different approach.  We will first choose a
convenient mass function, and then {\it  require\/} that the {\it two-body\/}
equation for a massles pseudoscalar bound state automatically have a 
solution when
the bare quark mass is zero.  In this case the quark mass function and the wave
function for the massless Goldstone boson will not be identical, but 
at least the
existence of the Goldstone boson in the chiral limit is assured.  We will {\it
define\/} the quark mass function of flavor $f$ by
\begin{equation}
m_f(p)\equiv m_f^0+c(m_f^0)\,f(p),\label{definemass}
\end{equation}
where $m_f^0$ is the current quark mass of flavor $f$, and $f(p)$ is a
universal function defined by
\begin{equation}
f(p)\equiv \frac{1}{|p^2|+\Lambda^2}.
\end{equation}
The function $c(m_f^0)$ can be thought of as a polynomial in powers of
$m_f^0$. This is the typical structure of the mass function which is usually
obtained from the solution of the one body equation.

The reason for
not solving the one body equation, in our case, is two fold. The first
problem is the difficulty of incorporating one gluon exchange into the
one body equation. Because of the on-shell constraint in the loop
momenta, the one gluon exchange interaction leads to an ultraviolet
divergence. The second problem is associated with our choice of infrared
regularization of the linear interaction. The infrared singularities
are regulated by the $P\cdot q$  term in the denominator of the 
linear interaction
Eq.~(\ref{nqtothe4}), and this would imply that the resultant
mass function is a function of two arguments, i.e. $m=m(p^2,{\bf p}^2)$.
This is unacceptable, and rather than  forsaking important
features of the model such as confinement and asymptotic  freedom, we choose to
model the quark mass functions.
 
The form (\ref{definemass}) guarantees that at large momenta, quark masses go
to their current quark mass values as dictated by asymptotic freedom.  In the
chiral limit the quark mass reduces to
\begin{equation}
m_{\chi}(p)=c(0)\,f(p) \label{mass1}
\end{equation}
We fix the constant $c(0)$ by requiring that the pion bound state equation,
using the mass function (\ref{mass1}), give a massles solution. This 
insures that a
massless pion exists in the chiral limit when $m_f^0=0$.  Next we 
{\it choose\/} a
value for the light current 	quark mass, $m_u^0=m_d^0$, and fix 
$c(m_u^0)$ so that
the two-body equation gives the correct value for the physical pion mass.
This also fixes the value of the on-shell quark mass away from the 
chiral limit.
Similarly, we {\it choose\/} $m_s^0$ and fix $c(m_s^0)$ by fitting 
the kaon mass. For
three flavors it is therefore sufficient to  have a function 
$c(m_f^0)$ which is a
polynomial of order 2 in $m_f^0$. As new flavors are introduced the 
order of the
polynomial accordingly can be increased.

To summarize, we have 6 mass parameters: $m^0_u, m^0_s, c(0), 
c(m_u^0),c(m_s^0)$, and
$\Lambda$.    In practice we fix $\Lambda$ at one GeV and {\it 
choose\/} the current quark
masses $m^0_u,$ and $m^0_s$ to be near the values expected by current 
theory.  We then
adjust the $c$'s to give the a zero mass pion in the chiral limit, 
and a real pion
and kaon with the observed masses.  This process is repeated for 
different values of
the current quark masses and the potential parameters $\sigma$ and $C$ until
satisfactory values for the constituent quark masses and the spectrum 
of excited pions is
obtained.  The final values of the parameters will be given in the 
next section.

Having outlined the features of the model, we now turn our attention to the
details of the  pseudoscalar bound state equation with spin.

\section{Pseudoscalar channel}
\label{chiral_limit}

The bound state vertex function  has the following structure
\be
\chi=\chi_{\rm color}\otimes\chi_{\rm flavor}\otimes\chi_{\rm spin}.
\ee
The color space vertex function is a Kronecker delta function, $\delta_{cd}$,
which reflects the color singlet nature of the bound state. The flavor space
vertex function is the matrix $\lambda^i_{fg}$ in $SU(3)$ matrix space,
which chooses the right flavor combination of the meson under consideration.
Indices $f,g$ refer to up down and strange quark entries ($u,d,s=1,2,3$) of
$\lambda^i$. For example, $[\lambda^+]_{ud}=[\lambda^+]_{12}$. For a general
meson type $i$, the bound state vertex function is
\be
\chi_{\alpha\beta, fg,cd}^{i}(k_1,k_2)\equiv
\delta_{cd}\,\lambda^i_{fg}\,\Gamma_{\alpha\beta}(k_1,k_2).
\ee
where $\alpha, \beta$ are Dirac indices (to be suppressed in the 
following discussion).
The most general form for the spin-space part of the vertex function 
for  pseudoscalar
mesons is
\begin{eqnarray}
\Gamma(k_1,k_2)=\gamma_5\bigg{\{}\Gamma_0+\ \! / \! \! \! \! P\,\Gamma_1+\
\! /
\! \! \! k\,\Gamma_2+[\ \! / \! \! \! k,\ \! / \! \! \! \!
P]\,\Gamma_3\bigg{\}}  \, ,\label{pseudoscalarmode}
\end{eqnarray}
where $\Gamma_i=\Gamma_i(k_1,k_2)$ are scalar functions.  The 
dominant contribution to the
bound  state vertex function comes from the first term of 
(\ref{pseudoscalarmode}),
\begin{equation}
\Gamma(k_1,k_2)\approx \gamma_5\,\Gamma_0(k_1,k_2) \, ,
\label{pseudoscalar0}
\end{equation}
This approximation, which is exact in the chiral limit when $P=0$ and 
$m_1=m_2$,
will be used for  the pion and kaon bound states in this work.

Assuming (\ref{pseudoscalar0}), multiplying the four channel equations for
pseudoscalar mesons by
$\gamma_5$, and taking the trace, gives the following approximate 
coupled equations
for pseudoscalar states
\begin{eqnarray}
\Gamma_i^{s}({\bf p},M_B)=&-&\frac{1}{2}\sum_{jr}\int
\frac{d^3k}{(2\pi)^3\,2E_j(k)}
\Biggl\{V_{ij}^{sr}(p,k)
\left[F_j(k_j^r)\Gamma_j^r({\bf k},M_B)  - F_i(p_i^s)\Gamma_i^s({\bf p},M_B)
\right]\nonumber\\
&+&6\delta_{ij}\delta_{sr}V_g(p-k) F_j(k_j^r) \Gamma_j^r({\bf k},M_B)
\Biggr\}  + 2 F_i(p_i^s)C \Gamma_i^s({\bf p},M_B) \, ,
\label{4channel3}
\end{eqnarray}
where the four channel wave functions $\Gamma_i^{s}({\bf p},M_B)$ are 
obtained from
$\Gamma_0$ as shown in Eq.~(\ref{fourvertex}), and
\begin{eqnarray}
F_1(k_1^r)\equiv&&\frac{m_1 m_2(k_2)+k^r_1\cdot
k_2}{m_2^2(k_2)-k_2^2}\nonumber\\
F_2(k_2^r)\equiv&&\frac{m_1(k_1)m_2 +k_1\cdot
k^r_2}{m^2_1(k_2)-k_1^2}\, ,
\label{newfk}
\end{eqnarray}
where $m_i(k_i^r)=m_i(-k_i^r)=m_i$.  For future reference we record 
the four-momentum
$q_{ij}^{rs}\equiv (p_1-k_1)_{ij}^{rs}$ exchanged between the two quarks.  This
depends on the initial and final channel.  The distinct cases are:
\begin{eqnarray}
q_{11}^{rs}&&=q_{22}^{-r,-s}=(rE(p)-sE(k),\;{\bf p}-{\bf k})\nonumber\\
q_{12}^{rs}&&=(rE(p)+sE(k)-M_B,\;{\bf p}-{\bf k})\nonumber\\
q_{21}^{rs}&&=(M_B-rE(p)-sE(k),\;{\bf p}-{\bf k})\, .
\label{allqs}
\end{eqnarray}
The solution of Eqs.~(\ref{4channel3}) for a realistic
choice of the parameters will be discussed in the next section.

Before turning to this discussion, look at the coupled equations in the chiral
limit, when $P=0$ and the dynamical quark masses are equal, so that
$m_1(k)=m_2(k)=m(k)$.  In this limit,
$k_1=-k_2$, and expanding to order $P\cdot k_1^r$ gives
\begin{equation}
F_1(k_1^r)=\frac{m(k_1^r)\, m(P-k_i^r)+k^r_1\cdot
P-k_i^{r2}}{m^2(P-k_i^r)-(P-k_i^r)^2} = \frac{1-2mm'}{2-4mm'}=\frac{1}{2}
=F_2(k_2^r) \, ,
\end{equation}
where $m\equiv m(\pm k_i^s)$ and $m'\equiv d m(\pm k)/d k^2 |_{(k^2=m^2)}$.
Hence, using charge conjugation symmetry (\ref{gammatrans}), the four coupled
equations (\ref{4channel3}) reduce to only {\it two\/} equations in the chiral
limit.  These coupled equations are
\begin{eqnarray}
\Gamma^+_\chi({\bf p},0)=&-&\int
\frac{d^3k}{(2\pi)^3\,2E(k)}
\Biggl\{V_+(p,k)
\left[\Gamma^+_\chi({\bf k},0)  - \Gamma^+_\chi({\bf p},0)
\right]\nonumber\\
&+&V_-(p,k)\left[\Gamma^-_\chi({\bf k},0)- \Gamma^+_\chi({\bf 
p},0)\right]  +6V_g(p-k)
\Gamma^+_\chi({\bf k},0)
\Biggr\}  + 2 C \Gamma^+_\chi({\bf p},0) \nonumber\\
\Gamma^-_\chi({\bf p},0)=&-&\int
\frac{d^3k}{(2\pi)^3\,2E(k)}
\Biggl\{V_+(p,k)
\left[\Gamma^-_\chi({\bf k},0)  - \Gamma^-_\chi({\bf p},0)
\right]\nonumber\\
&+&V_-(p,k)\left[\Gamma^+_\chi({\bf k},0) - \Gamma^-_\chi({\bf 
p},0)\right]  +6V_g(p-k)
\Gamma^-_\chi({\bf k},0)
\Biggr\}  + 2 C \Gamma^-_\chi({\bf p},0) \, ,
\label{cheq}
\end{eqnarray}
where
\begin{eqnarray}
V_\pm(p,k) = \frac{8\pi\sigma}{({\bf p}-{\bf k})^4 + (E(p)\mp E(k))^4} \, .
\label{cheq2}
\end{eqnarray}
Note that these two equations are symmetric under the interchange
\be
\Gamma^+_\chi\leftrightarrow\pm\Gamma^-_\chi \, , \label{symmetry}
\ee
and hence reduce to one equation for 
$\Gamma_\chi\equiv\Gamma^+_\chi=\pm\Gamma^-_\chi$
\begin{eqnarray}
\Gamma_\chi({\bf p},0)=&-&\int
\frac{d^3k}{(2\pi)^3\,2E(k)}
\Biggl\{\left[V_+(p,k)\pm V_-(p,k)  +6V_g(p-k) \right]
\Gamma_\chi({\bf k},0) \nonumber\\
&-& \left[V_+(p,k) + V_-(p,k) \right] \Gamma_\chi({\bf p},0) \Biggr\}  + 2 C
\Gamma_\chi({\bf p},0) \, ,
\label{cheq3}
\end{eqnarray}
where the sign of the $V_-$ term depends on the sign in the relation 
(\ref{symmetry}).
Since the $\pi_0$ is even under charge conjugation symmetry, the plus 
sign is the
correct one to use.

Recalling Eq.~(\ref{mass1}), the energies $E$ in Eq.~(\ref{cheq3}) 
depend on the
constant $c(0)$
\begin{eqnarray}
E(p) = \sqrt{c(0)^2f^2(p)+{\bf p}^2} \, ,
\label{cheq2a}
\end{eqnarray}
and this is adjusted to insure that the Eq.~(\ref{cheq3}) has a solution.  Once
$c(0)$ has been fixed, Eqs.~(\ref{4channel3}) are solved for various values of
the bare quark masses $m_f^0$ and the ``mass functions'' $c(m_f^0)$, and all
parameters are adjusted to give a reasonable spectrum.

Having outlined the features of the model we next present the results for
mass functions of quarks and vertex functions for bound states.

\begin{figure}
\begin{center}
\mbox{
    \epsfxsize=4.0in
\epsffile{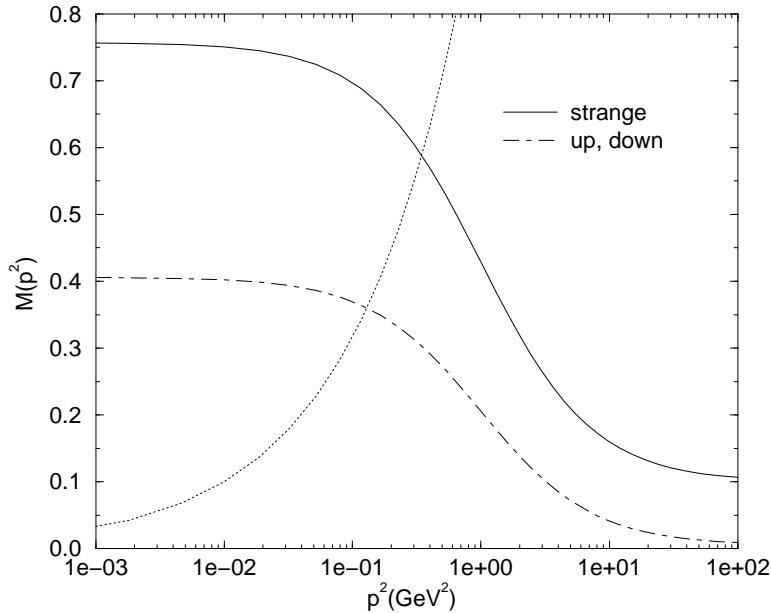}
}
\end{center}
\caption{Quark mass functions $m_f(p)\equiv M(p^2)$ are shown for 
up/down, and strange
quarks. On-shell quark masses are $m_{u,d}=360$ MeV, and $m_s=588$ MeV. At
large momenta quark mass values approach to $m_{u,d}^0=5$ MeV, and $m_s^0=100$
MeV. }
\label{mums.fig}
\end{figure}

\begin{table}
\begin{center}
\begin{minipage}{4.0in}
\caption{Summary of results}
\begin{tabular}{ccc}
Observable & Calculated & Experimental \\ \hline\hline
$m_{\pi}$  & 140 MeV & 139.6 MeV \\
$m_{\eta}$  & 320 MeV & ---  \\
$m^*_{\pi}$   & 1118 MeV & 1300$\pm$ 100 MeV\\
$m_{K}$  & 495 MeV & 495 MeV\\
$m_\chi$   &  376 MeV  & --- \\
$m_u=m_d$  &  360 MeV & --- \\
$m_s$   &  588 MeV  & --- \\
\end{tabular}
\label{table}
\end{minipage}
\end{center}
\end{table}

\section{Results}
\label{results}

The quark mass functions are shown in Fig.~\ref{mums.fig}. The on-shell quark
masses $m_f$ are given in Table I. At  large momenta, the quark mass
values approach the bare quark masses $m_f^0$ shown in Table II.  The 
other mass
parameters and bound state parameters are also shown in Table II. The parameter
$\Lambda$ which determines the scale of mass function was fixed at 
$\Lambda=1$ GeV
and not adjusted during the fits. The third line in 
Fig.~\ref{mums.fig} is the momentum
$p$, and the intersection of this line with the quark mass function gives the
constituent quark mass.
 
\begin{figure}
\begin{center}
\mbox{
    \epsfxsize=4.0in
\epsffile{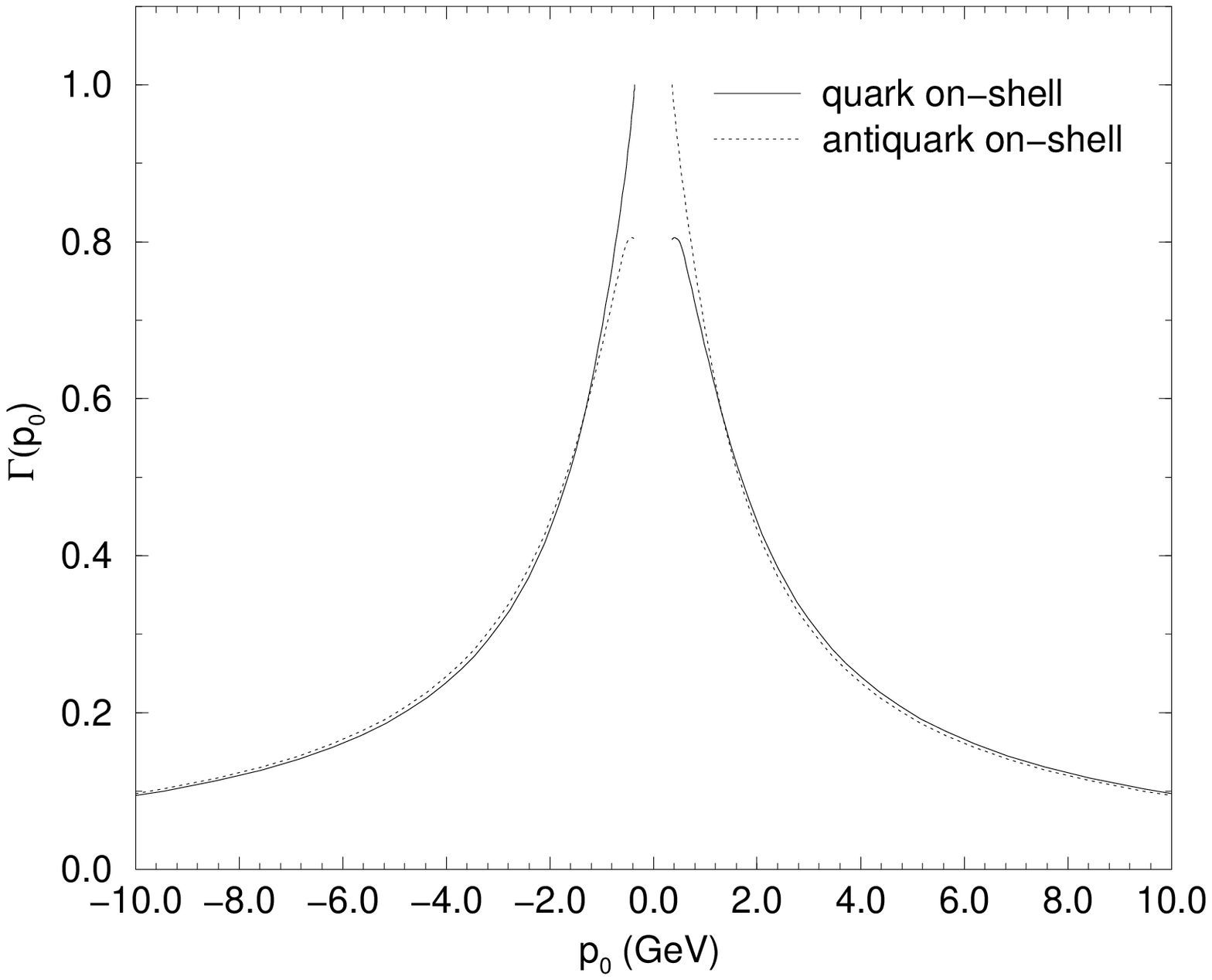}
}
\end{center}
\caption{The four-channel vertex functions for the ground state of the pion.}
\label{pi.wfn.plt}
\end{figure}

\begin{figure}
\begin{center}
\mbox{
    \epsfxsize=4.0in
\epsffile{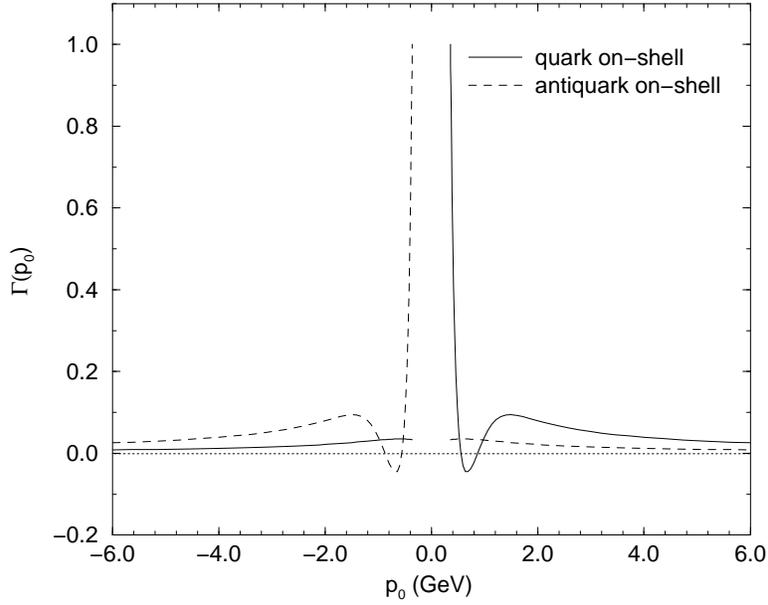}
}
\end{center}
\caption{The four-channel vertex functions for the first excited state of the
pion.}
\label{pistar1.wfn.plt}
\end{figure}

\begin{figure}
\begin{center}
\mbox{
    \epsfxsize=4.0in
\epsffile{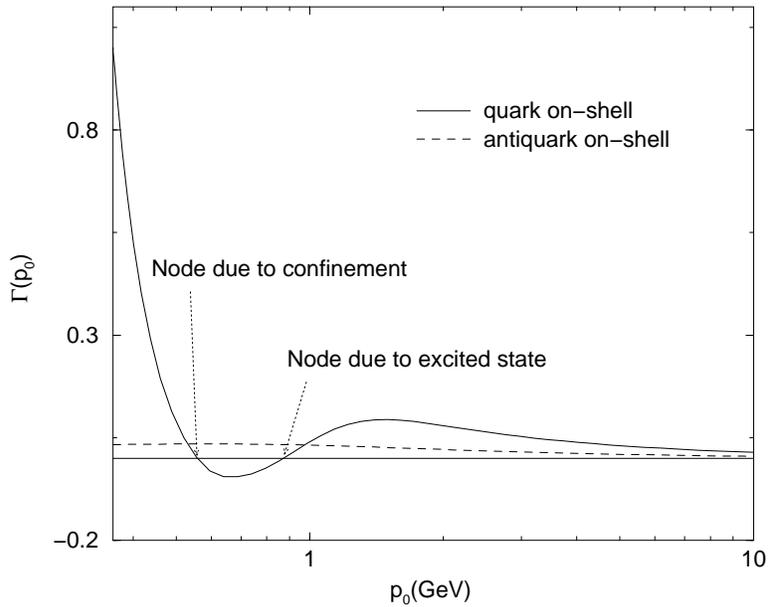}
}
\end{center}
\caption{The two positive energy vertex functions for the first 
excited state of
the pion. The second node is due to the excited state, and the first 
node assures
that the bound state does not decay.}
\label{pistar2.wfn.plt}
\end{figure}

In Figs.~\ref{pi.wfn.plt} and ~\ref{pistar1.wfn.plt} the ground and 
first excited
state vertex functions of the pion are shown.  Here we show the 
vertex functions as a
function of the variable $p^s_j= s E(p)\equiv s p_0$.  Note that $p_0$ is
positive for positive energy states ($s=+$) and negative for negative 
energy states
($s=-$).  Because of the symmetrization, the positive energy quark 
vertex function is the
same as the negative energy anti-quark vertex function up to an 
overall phase ($+$ for
states even under charge conjugation and $-$ for odd states). Also 
note that the curves are
not continuous because the argument $p_0$ can not take  values between 
$(-m, +m)$. In
Fig.~\ref{pistar2.wfn.plt} we present the excited state vertex 
functions on a logarithmic
scale.  The location of the first node is exactly  where both quarks 
are simultaneously on
shell. Therefore, although  kinematically allowed, the excited state 
of the pion can not
decay into a free  quark-antiquark pair. {\it This numerical result 
is a consequence of the
confinement condition (\ref{confcond})\/}.

\begin{figure}
\begin{center}
\mbox{
    \epsfxsize=4.0in
\epsffile{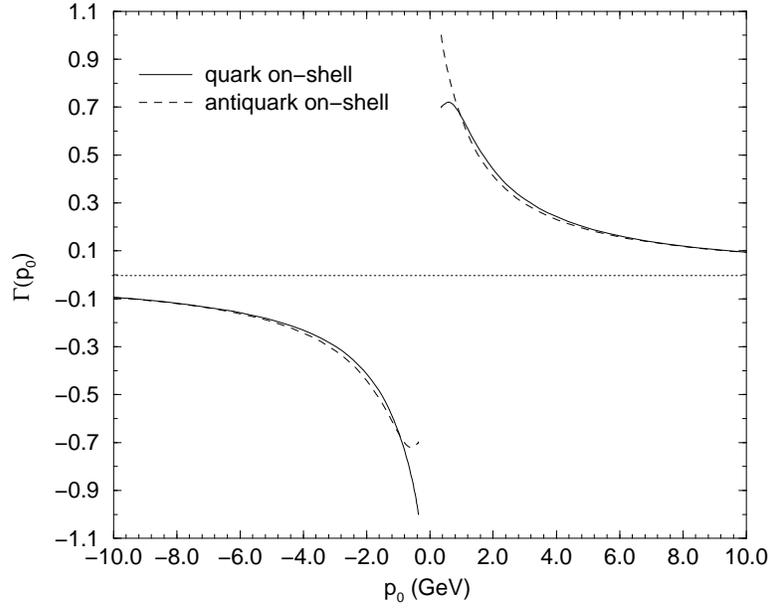}
}
\end{center}
\caption{The four-channel vertex functions for the non-strange $\eta$.}
\label{pi.wfnlog.plt}
\end{figure}

\begin{figure}
\begin{center}
\mbox{
    \epsfxsize=4.0in
\epsffile{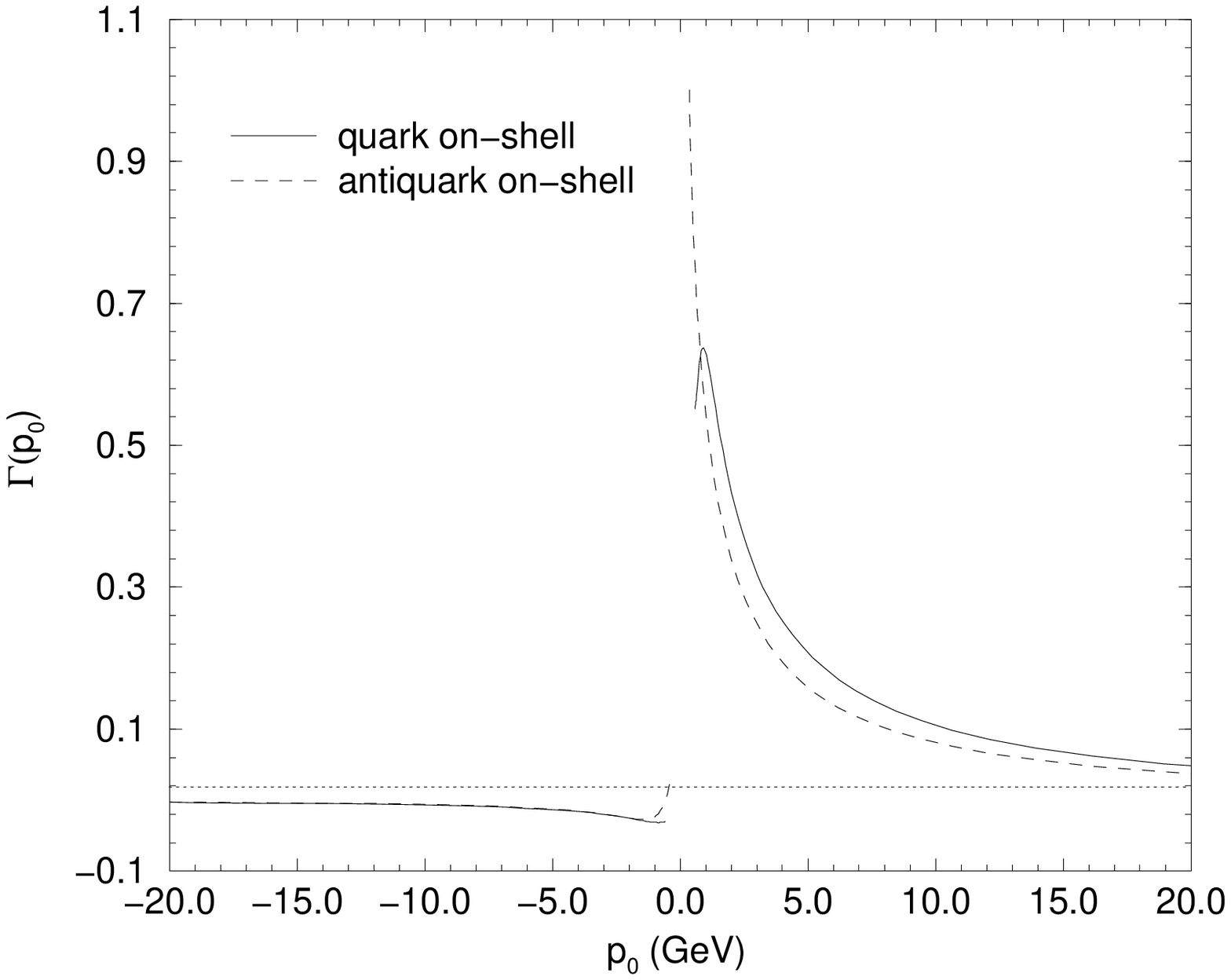}
}
\end{center}
\caption{The four-channel vertex functions for the ground state of the kaon.}
\label{kaon.wfn.plt}
\end{figure}

In Fig.~\ref{pi.wfnlog.plt} we present the non strange-eta (the isospin zero
$u\bar{u}+d\bar{d}$ combination) ground state vertex functions.  Note 
that these are
odd under charge conjugation.  The kaon vertex functions are shown in
Fig.~\ref{kaon.wfn.plt}.  Since the kaon is formed from a quark and antiquark
of unequal masses, the particle-antiparticle symmetry is lost and the 
negative and
positive energy solutions have a differerent shape and size.

\begin{figure}
\begin{center}
\mbox{
    \epsfxsize=4.0in
\epsffile{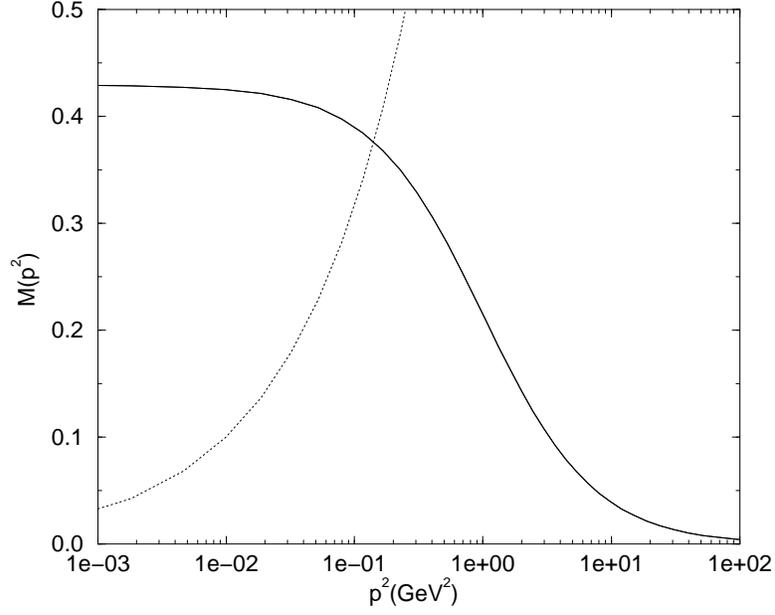}
}
\end{center}
\caption{The chiral limit of the quark mass function $M(p^2)\equiv 
m_{\chi}(p)$. The
on-shell  quark mass is $m_{\chi}=376$ MeV. At large momenta quark 
mass function
approaches 0.}
\label{mumsch.fig}
\end{figure}

\begin{figure}
\begin{center}
\mbox{
    \epsfxsize=4.0in
\epsffile{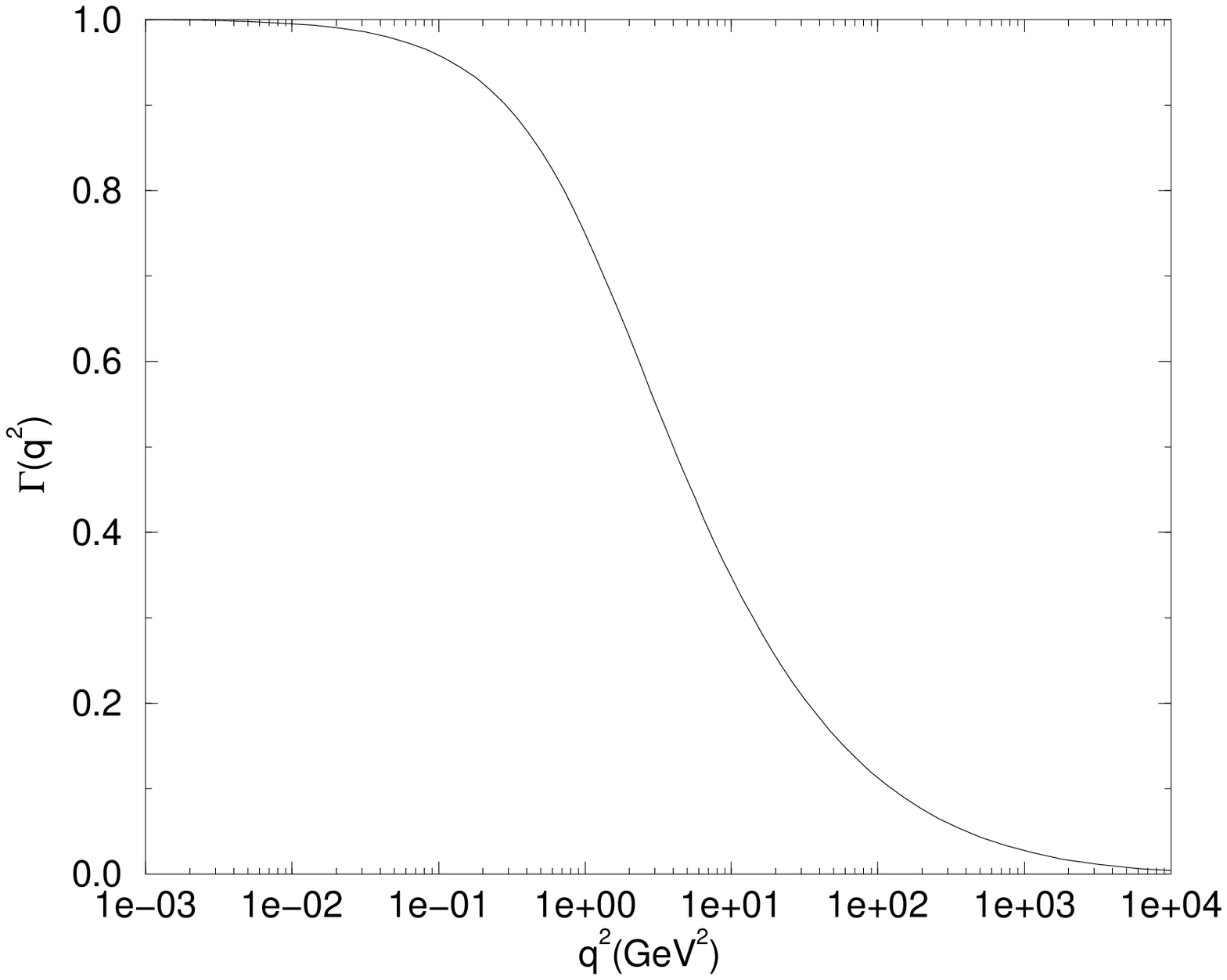}
}
\end{center}
\caption{The chiral limit of the pion ground state vertex function.}
\label{pich.wfn.plt}
\end{figure}

The mass function and the pion wave function in the chiral limit are shown in
Figs.~\ref{mumsch.fig} and \ref{pich.wfn.plt}.

\begin{table}
\begin{center}
\begin{minipage}{2.5in}
\caption{Values of the parameters}
\begin{tabular}{cc}
Parameter & Value \\ \hline\hline
$m_u^0$  & 5 MeV \\
$m_s^0$   & 100 MeV  \\
$c(0)$  & 0.429 (Gev)$^3$ \\
$c(m_u^0)$  &  0.400 (GeV)$^3$\\
$c(m_s^0)$  &  0.657 (GeV)$^3$ \\
$\sigma$  &  0.4 (GeV)$^2$ \\
$C$  &  $0.4929 $  \\
$\Lambda$  &  1 GeV \\
\end{tabular}
\label{table2}
\end{minipage}
\end{center}
\end{table}


\section{CONCLUSION}
\label{conclusion}

We have shown that a relativistic generalization of the Schr\"odinger equation
with linear interaction leads to the Gross equation. It is not possible to
write a Bethe Salpeter equation that gives the correct linear 
interaction in the
nonrelativistic limit. We have proved that  the relativistic 
generalization of the linear
interaction leads to vanishing  vertex amplitudes when both of the 
constituents are
on-shell. This guarantees  that the bound state does not decay to its 
constituents.
This mechanism of confinement follows from insisting on the correct 
nonrelativistic limit.
The model incorporates asymptotic freedom through the inclusion of a 
vector one gluon
exchange interaction, and quark mass functions that approach the 
current quark values at
infinite momentum. There are no cut-off's or  ad-hoc form factors 
involved, and the linear
interaction involves only one coupling parameter.  The approach give a 
good description of
the pion, kaon, and eta.

It remains to use this formalism to describe the full meson spectrum.

\section{Acknowledgements}
This work was supported in part by the US Department
of Energy under grant No.~DE-FG02-97ER41032. One of us (\c{C}\c{S}) would 
like to thank the theory group of the Jefferson Lab for the hospitality, where
part of this work was completed.

\label{section5}
\appendix
\section{Numerical methods}

Solutions of integral equations are performed by first discretizing the
integrals
\be
\int dq\, f(q) \longrightarrow \sum_{i=1}^n w_i\, f(q_i),
\ee
where $w_i$ are integration weights for grid points $q_i$. In order to map
the grid points and weights from interval $(-1,1)$ to $(0,\infty)$ we use the
arctangent mapping (Ref.~\cite{THK,TK})
\be
y(x)=R_{min}+\frac{R_{d}\;{\displaystyle{\rm
tan} \,\left(\frac{\pi}{4}(1+x)\right)}}{\displaystyle 
{1+\frac{R_{d}}{R_{max}-R_{min}}{\rm
tan}\,\left(\frac{\pi}{4}(1+x)\right)}}
\, ,
\label{mapping}
\ee
where
\be
R_{d}=\frac{R_{med}-R_{min}}{R_{max}-R_{med}}(R_{max}-R_{min}).
\ee
It follows that
\be
y(-1)=R_{min}\,\qquad y(0)=R_{med}\, ,\qquad y(1)=R_{max}\, .
\ee
Therefore, one can safely control the range $(R_{min},R_{max})$ and 
distribution
$(R_{med})$ of grid  points. With this discretization procedure, 
continuous  integral
equations are transformed into nonsingular matrix equations.

The spectator equation is an eigenvalue problem, where the eigenvalue is
the mass of the bound state. The equation can be brought into the following
form
\be
\sum_{j=1}^{N}[H_{M}(p_i,p_j)-1]\Phi(p_j)=0.
\label{matrix}
\ee
where $M$ is the bound state mass, and $p_i\,(i=1...N)$ are grid points
in momentum space. Therefore, $H$ is an $N\times N$ matrix and $\Phi$ is a
vector of dimension $N$, which leads to the following matrix equation
\be
[H_{M}-1]\Phi=0,
\label{matrixeq}
\ee
where $M$ is unknown. Start by making an initial guess for $M$. In order
to find the ground state, one should start with an initial guess near the
expected value of the ground state mass. The next step is to see whether the
initial guess leads to a consistent solution. The most efficient way of
checking whether a given matrix has a specific eigenvalue is through the
method of inverse iteration, as suggested in Refs.~\cite{THK,TK}. 
First construct an
arbitrary vector $\chi^0$
\be
\chi^0=\sum_{i=1}^N c_i\,\Phi_i
\ee
where $\Phi_i$, $i=1..N$, satisfy
\be
[H_{M}-\omega_i]\Phi_i=0,
\ee
where $\omega_i,\,i=1..N$ are eigenvalues of the $H_{M}$ matrix. It should
be emphasized that eigenvalues which are not equal to 1
have no physical meaning, for they do not correspond to a solution of the
equation (Eq.~\ref{matrixeq}). Next, construct
\be
K=\frac{1}{H_{M}-1}\, .
\ee
Letting $K$ operate on state $\chi^0$ $n$ times produces
\be
\chi^n=K^n\,\chi^0=\sum_{i=1}^N\,\frac{c_i}{(\omega_i-1)^n}\,\Phi_i.
\ee
When the number of iterations $n$ is sufficiently large (usually around ten),
the dominant contribution to $\chi^n$ comes from the eigenvector $\Phi_j$
whose eigenvalue $\omega_j$ satisfies $|\omega_j-1|<|\omega_i-1|$
for all $i=1 \cdots j-1,j+1 \cdots N$. Therefore,
\bea
\chi^{n}&\approx& \frac{c_j}{(\omega_j-1)^n}\,\Phi_j,\nonumber\\
\chi^{n+1}&\approx& \frac{1}{\omega_j-1}\,\chi^n.
\eea
Using the eigenvector $\chi^n$, which is proportional to $\Phi_j$, 
the eigenvalue
$\omega_j$ can be found from
\be
\omega_j=\frac{\chi^{n\dagger}H\chi^n}{\chi^{n\dagger}\chi^n}.
\ee
If $\omega_j$ is close enough to 1, then one has a self consistent
solution. This method has the benefit of directly singling out the eigenvalue
closest to the initial guess, rather than finding the largest eigenvalue as in
the case of straight forward iteration. Excited states can similarly be found
by varying the initial guess $M$ towards higher values. There is only one
matrix inversion involved. Distribution of the grid
points in momentum space is done by the arctangent mapping. The typical
number of momentum space grid points used in order to obtain stable solutions
is around 40.


\begin{references}
\bibitem{GROSS} F.~Gross and J.~Milana, Phys. Rev. D {\bf 43}, 2401 (1991);
{\bf 45}, 969 (1992); {\bf 50}, 3332 (1994).
\bibitem{ROBERTS1} C. D. Roberts and A. G. Williams, Prog. Part. Nucl. Phys.
{\bf 33}, 477 (1994).
\bibitem{JM92} R.~L.~Jaffe and P.~F.~Mende, Nucl.~Phys.~{\bf B369}, 189 (1992).
\bibitem{O97} M.~G.~Olsson, Phys. Rev. D {\bf 56}, 283 (1997).
\bibitem{grossref} F.~Gross, Phys.~Rev.~{\bf 186}, 1448 (1969).
\bibitem{GROSS2} J.~Adam, J.~W.~Van Orden, and F.~Gross 
Nucl.~Phys.~{\bf A640}, 391
(1998).
\bibitem{cc} F. Gross, preprint JLAB-THY-99-24, WM-99-115, nucl-th/9908084
\bibitem{GVOH} F.~Gross, J.~W.~Van Orden, and K.~Holinde, 
Phys.~Rev.~C {\bf 45}, 2094
(1992).
\bibitem{AGVO} J.~Adam, F.~Gross, \c{C}.~\c{S}avkl{\i}, and J.~W.~Van Orden,
Phys.~Rev.~C {\bf 56}, 641 (1997).
\bibitem{CETIN} \c{C}.~\c{S}avkli and F.~Tabakin, Nucl. Phys. A 628 
(1998) 645-668.
\bibitem{THK} D. Heddle Y. R. Kwon F. Tabakin, Comp. Phys. Comm. {\bf 38}
(1985) 71.
\bibitem{TK} Y. R. Kwon F. Tabakin, Phys. Rev. C {\bf 18} (1978) 932.

\end{references}
\end{document}